\documentclass[12pt]{iopart}

\usepackage{graphicx}
\usepackage{epsfig}
\usepackage[english]{babel}

\newcommand{\be}{\begin{equation}}
\newcommand{\bea}{\begin{eqnarray}}
\newcommand{\eea}{\end{eqnarray}}
\newcommand{\ee}{\end{equation}}

\newcommand{\HH}{\mathcal{H}}
\newcommand{\HHk}{\mathcal{H}^{\rm k}}

\def\one{\ensuremath{\hbox{$\mathrm I$\kern-.6em$\mathrm 1$}}}
\def\tr{ \mbox{tr}}
\def\qed{\leavevmode\unskip\penalty9999 \hbox{}\nobreak\hfill
     \quad\hbox{\leavevmode  \hbox to.77778em{%
               \hfil\vrule   \vbox to.675em%
               {\hrule width.6em\vfil\hrule}\vrule\hfil}}
     \par\vskip3pt}
\newcommand{\beaa}{\begin{eqnarray*}}
\newcommand{\eeaa}{\end{eqnarray*}}
\newcommand{\bma}{\begin{subequations}}
\newcommand{\ema}{\end{subequations}}

\def\S{{\cal S}}

\def\one{{\bf 1}}

\def\noxrightarrow[#1]{\dodoublegroupempty\dodoxrightarrow{#1}}
\def\noxleftarrow [#1]{\dodoublegroupempty\dodoxleftarrow {#1}}
\def\dodoxrightarrow#1#2{\mathrel{{\domthxarr0359\rightarrowfill{#1}{#2}}}}
\def\dodoxleftarrow#1#2{\mathrel{{\domthxarr3095\leftarrowfill{#1}{#2}}}}

\usepackage{iopams}

\begin{document}

\title{Renormalization and tensor product states in spin chains
and lattices}

\author{J. Ignacio Cirac}
\address{Max-Planck-Institut f\"ur Quantenoptik,
Hans-Kopfermann-Str. 1, D-85748 Garching, Germany.}
\ead{ignacio.cirac@mpq.mpg.de}

\author{Frank Verstraete}
\address{Fakult\"at f\"ur Physik, Universit\"at Wien,
Boltzmanngasse 5, A-1090 Wien, Austria.}
\ead{frank.verstraete@univie.ac.at}

\begin{abstract}
We review different descriptions of many--body quantum systems in
terms of tensor product states. We introduce several families
of such states in terms of known renormalization procedures,
and show that they naturally arise in that context. We concentrate on
Matrix Product States, Tree Tensor States, Multiscale Entanglement
Renormalization Ansatz, and Projected Entangled
Pair States. We highlight some of their properties, and show how
they can be used to describe a variety of systems.
\end{abstract}


\maketitle

\section{Introduction}

Many-body quantum states appear in many contexts in Physics and other
areas of Science. They are very hard to describe, even computationally,
due to the number of parameters required to express them, which typically
grows exponentially with the number of particles. Let us consider a
spin chain of $N$ spin $s$ systems. If we write an arbitrary state
in the basis $|n_1,\ldots,n_N\rangle$, where $n_k=1,\ldots, (2s+1)$,
we will have to specify $(2s+1)^N$ coefficients. Even if $s=1/2$, and
$N\sim 50$, it is impossible to store such a number of coefficients
in a computer. Furthermore, even if that would be possible, whenever
we want to make any prediction, like the expectation value of an observable,
we will have to operate with those coefficients, and thus the number of
operations (and therefore the computational time) will inevitably grow
exponentially with $N$.

In many important situations, one can circumvent this problem by using
certain approximations. For example, sometimes it is possible to describe
the state in the so-called mean field approximation, where we write
$|\Psi\rangle=|\phi_1,\ldots,\phi_N\rangle$, ie. as a product state.
Here, we just have to specify each of the $|\phi_M\rangle$, and thus
we need only $(2s+1)N$ coefficients. This method and its extensions,
even though it has a restricted validity in general, has been very
successfully used to describe many of the phenomena that appear in
quantum many-body systems. This indicates that, among all possible states,
the ones that are relevant for many practical situations possess the
same properties as product states. Another very
successful method is renormalization \cite{wilson75,kadanoff}
which, in the context of spin chains and lattices,
tries to obtain the physics of the low energy states by grouping degrees
of freedom and defining new ones that are simple to handle, so that
at the end we can cope with very large systems and using very few parameters.
There exist very successful methods to uncover the physical properties
of many-body systems which exploit numerical approaches. The first one is
Quantum Monte Carlo, which samples product states in order
to get the expectation values of physical observables. Another one is
Density Matrix Renormalization Group (DMRG) \cite{white92,white92b}, which
is specially suited for 1D lattices, and is based on renormalization
group ideas.

\subsection{DMRG and Tensor Product States}

Wilson's renormalization
method provided a practical way of qualitatively determining the low energy behavior
of some of those systems. However, it was by no means sufficient to describe them
quantitatively. In 1991, Steve White \cite{white92,white92b} proposed a new way of performing the
renormalization procedure in 1D systems, which gave extraordinary precise results.
He developed the DMRG algorithm, in which
the renormalization procedure takes explicitly into account the whole system at each
step. This is done by keeping the states of subsystems which are relevant
to describe the whole wavefunction, and not those that minimize the energy on that
subsystem. The algorithm was rapidly extended and adapted to different situations
\cite{peschel,schollwoeck04}, becoming the method of choice for 1D systems.
In 1995, Ostlund and Rommer \cite{oestlund95} realized that the state
resulting from the DMRG algorithm could be written as a so--called Matrix Product State (MPS),
ie, in terms of products of certain matrices (see also
\cite{dukelsky98,takasaki99,verstraeteporras04}). They proposed to use
this set of states as a variational family for infinite homogeneous systems, where
one could state the problem without the language of DMRG,
although the results did not look as precise as with the finite version of that method.
Those states had appeared in
the literature in many different contexts and with different names.
First, as a variational Ansatz for the
transfer matrix in the estimation of the partition function of a classical
model \cite{Kramers}. Later on, in the AKLT model in 1D \cite{AKLT1,AKLT2}, where
the ground state has the form of a valence bond solid (VBS) which can be
exactly written as a MPS. Translationally invariant
MPS in infinite chains were thoroughly studied and characterized mathematically in full
generality in Ref. \cite{fannes92}, where they called such family
{\it finitely correlated states} (FCS). The name MPS was coined later on by
Kl\"umper et al \cite{kluemper93,kluemper94}, who introduced different
models extending the AKLT where the ground state had the explicit FCS form.
All those studies were carried out for translationally invariant systems,
where the matrices associated to each spin do not depend on the position of the spin.
An extension of the work of Fannes {\it et al} \cite{fannes92}
to general MPS (ie finite and non--homogeneous
states) appeared much later on \cite{perezgarcia06}.

Given the success of DMRG in 1D, several authors tried to extend it to higher
dimensions. The first attempts considered a 2D system as a chain, and used
DMRG directly on the chain \cite{liang94,white96,xiang01}, obtaining much less
precise results than in 1D. Another attempt considered homogeneous 3D classical system and
used ideas taken from DMRG to estimate the partition function of the
Ising Model \cite{CTMRG3D}. A different approach was first suggested by
Sierra and Martin-Delgado \cite{Sierra} inspired by the ideas of Ostlund and
Rommer \cite{oestlund95}. They introduced two families of translationally invariant states,
the {\it vertex-} and {\it face- matrix product state Ans\"atze}, and proposed
to use them variationally for 2D systems, in the same way as Ostlund
and Rommer used FCS in 1D. The first family generalized the AKLT 2D VBS state \cite{AKLT1}
in as much the same way as FCS did it in 1D. The second one was
inspired by interaction-round-the-face models in (classical) Statistical Mechanics.
The inclusion of few parameters in the VBS wavefunction of AKLT to extend that model
was first suggested in Ref. \cite{VBS1}. The authors also showed that the calculation
of expectation  values in those VBS could be thought of as evaluating a classical partition
function, something they did using Monte Carlo methods \cite{VBS1}, and Hieida {\it et al}
using ideas taken from DMRG \cite{VBS2}. Later on, Nishino and collaborators used
the representations proposed by Sierra and Martin-Delgado, as well as another one they
called {\it interaction-round-a-face} (inspired by the specific structure of the
transfer matrix of the classical Ising model) to determine the partition function of the
classical Ising Model in 3D variationally \cite{Ni1,Ni2,Ni4,Ni5,Ni6}.
For instance, in \cite{Ni5} a vertical density matrix algorithm was introduced to
calculate thermodynamical properties of that model based on the
the interaction-round-a-face representation and in \cite{Ni4} a perturbation approach
was taken for the vertex matrix product state, which turn out to be numerically unstable.
Eventually, quantum systems at zero temperature were considered by direct
minimization of trial wavefunctions of the VBS type with few variational parameters
\cite{SBS,IRF1}. In summary, most of the attempts in 2D quantum systems (and 3D classical ones)
tried to generalize the method of Ostlund and Rommer \cite{oestlund95} by using
families of states that extended FCS to higher dimensions, and dealing with
infinite homogenous systems. FCS and their extensions were based on tensors
contracted in some special ways, and thus all of them were called {\it tensor
product states} (TPS). Nevertheless, no DMRG--like algorithm for 2D or higher dimensions
was put forward (except for the direct application of DMRG by considering the 2D
system as a 1D chain).

The success of DMRG for 1D systems indicated that the family of
states on which it based, namely MPS, may provide an efficient and accurate description
of spin chain systems. In higher dimensions, however, the situation was
much less clear since only infinite systems were considered and the numerical
results were not entirely satisfactory.

\subsection{The corner of Hilbert space}

One can look at the problem of describing many--body quantum systems from a
different perspective.
The fact that product states in some occasions may capture the physics
of a many-body problem may look very surprising at
first sight: if we choose a random state in the Hilbert space (say, according
to the Haar measure) the overlap with a
product state will be exponentially small with $N$. This apparent contradiction
is resolved by the fact that the states that appear in Nature are not random states,
but they have very peculiar forms. This is so because of the following reason. If
we consider states in thermal equilibrium, each state of a system, described by
the density operator $\rho$, is completely
characterized by the Hamiltonian describing that system, $H$, and the temperature, $T$,
$\rho\propto e^{-H/T}$. In all systems we know, the Hamiltonian contains terms with at most
$k$--body interactions, where $k$ is a fixed number independent of $N$ which typically
equals 2. We can thus parameterize all possible Hamiltonians in Nature in terms of
$(N,k)\times (2s+1)^{2k}$. The first term is the number of groups of $k$ spins,
whereas the second one gives the number of parameters of a general Hamiltonian acting
on $k$ spins. This number scales only polynomially with $N$, and thus all possible
density operators will also depend on a polynomial number of parameters. In practice,
if we just have 2--body interactions (ie, k=2), and short--range interactions, the
number will be linear in $N$. If we additionally have translational symmetry, the number
will even be independent of $N$. This shows that even though we just need an
exponential number of parameters to describe a general state, we need very
few to describe the relevant states that appear in Nature. In this sense,
the relevant states are contained in "a corner of the Hilbert space". This representation
is, however, not satisfactory, since it does not allow one to calculate
expectation values.

These facts define a new challenge in many-body quantum physics, namely, to find good
and economic descriptions of that corner of Hilbert space. That is, a family
of states depending on few parameters (which increase only polynomially with $N$),
such that all relevant states in Nature can be approximated by members of such
family. If we are able to do that, as well as to characterize and study the properties
of such family, we would have a new language to describe many-body quantum systems
which may be more appropriate than the one we use based on Hilbert space expansions.
Apart from that, if we find algorithms which, for any given problem (say, a Hamiltonian
and a temperature), allows us to determine the state in the family which approaches
the exact one, we will have a very powerful numerical method to describe quantum many-body
systems. It is clear that product states are not enough for that task, since they
do not posses correlations (nor entanglement), something that is crucial in many physical
phenomena. So, the question is how to extend product states in a way that they
cover the relevant corner of Hilbert space.

One possible strategy to follow is to determine a property that all the states on
that corner have, at least for a set of important problems. If we then find a family
of states which includes all states
with that particular property, we will have succeeded in our challenge. But, what
property could that be? Here, the answer may come from ideas developed in the
context, among others, of quantum information theory. One of such ideas appears
for the subclass of problems with Hamiltonians that contain finite-range interactions
(ie, two spins interact if they are at a distance smaller than some constant),
have a gap (ie, for all $N$, $E\ge E_0 +\Delta$, where $E$ is the energy of any excited state,
$E_0$ that of the ground state, and $\Delta>0$), and are at zero temperature. In
that case, the so--called area law naturally emerges
\cite{Srendnicki,Wilczek,Wilczek2,vidallatorre03,calabrese,areareview}. It states that for the ground
state of such a Hamiltonian, $|\Psi_0\rangle$, if we consider a block, $A$, of
neighboring spins, the von Neumann entropy of the reduced density operator of such a region,
$\rho_A$, scales with the number of particles at the border of that region. This
is quite remarkable, since the von Neumann entropy being an extensive quantity, for
a random state it will scale with the number of particles in $A$, and not in the
border. The area law has been proven in 1D spin chains \cite{Hastingsarea}, and it is fulfilled for
all Hamiltonians we know in higher dimensions. Even for critical systems, where
the gap condition is not fulfilled, only a slight violation occurs (namely that
it is proportional to the number of spins at the border, $L$, times $\log L$)
\cite{calabrese,wolf06,gioev06}.
What happens at finite temperature? In that case one can also find a global
property of all states in the corner of Hilbert space which extends the area
law. In contrast to the zero temperature case, now this can be rigorously proven for
any Hamiltonian possessing finite-range interactions in arbitrary
dimensions \cite{wolfmutual07}. The property is the following:
given a region $A$ as before, the mutual information between the spins in that region
and the rest of the spins (in region $B$) is bounded by a constant times the number of spins
at the border divided by the temperature. Here, the mutual information is defined
as $I(A:B)=S(\rho_A)+S(\rho_B)-S(\rho_{AB})$, where $S$ is the von Neumann entropy, and
$\rho_X$ is the density operator corresponding to region $X$.

There is a way of constructing families of states explicitly fulfilling the area law.
Let us first consider a 1D chain with two spins per site in which we entangle each of them
with the nearest neighbor spins (to the right and to the left, respectively). If we
now consider a block of neighboring sites, only the outermost spins with contribute
to the entropy of that block, and thus the area law will be fulfilled. Furthermore,
if we project in each site the state of the two spins onto a subspace of lower
dimension, the resulting state will also fulfill the area law. This construction
can be straightforwardly extended to higher dimensions, just by replacing each spin
in the lattice by $z$ auxiliary spins which are pair--wise entangled with their
neighbors (here $z$ is the coordination number of the lattice). By projecting
the state of the auxiliary spins onto the Hilbert space of the original spin at each
site, we obtain a state automatically fulfilling the area law. This construction, which
is inspired by the VBS \cite{AKLT1}, was introduced in the context of quantum
information in order to study localizable entanglement of spin chains \cite{localizable} and to
give an alternative explanation of the measurement based model of quantum
computation \cite{verstraetecirac04b}. Later on, it was extended to describe general
spin lattice systems in \cite{verstraetecirac04}, where the resulting states
where called {\it Projected Entangled-Pair States} (PEPS), given the way they were
defined. In that work, a method to approximate the ground state wavefunction
in terms of PEPS for finite systems was introduced, which allowed to find the
optimal projectors at each site in very much the same way as DMRG does it in 1D.
Thus, the method can be considered as a truly extension of the DMRG algorithm to
higher dimensions, although computationally it is much more demanding. The key
ingredient in that method was a new algorithm that
allows one to optimally approximate arbitrary states by PEPS in a variational fashion.

By explicitly expressing the projectors appearing in the PEPS in an orthonormal basis,
one can immediately see that they have a tensor product form. In fact, in 1D
the family of PEPS coincides with the MPS \cite{localizable}. In higher dimensions,
if one takes an infinite system and chooses all the projectors to be identical, the
so-called iPEPS \cite{jordan07}
coincide with the vertex matrix product ansatz introduced by Sierra and Martin-Delgado
\cite{Sierra}. The PEPS construction, however, apart from giving rise to (finite) DMRG--like
algorithms in higher dimensions, gives a clear picture of the entanglement properties
of those states (as, eg, related to the area law) as well as how
correlations are carried over by the entangled auxiliary particles.
More importantly, it can be used to
extend the PEPS to fermionic systems still keeping all the properties that
made them special \cite{KrausFermions}.

The fact that PEPS fulfill the area law puts them in a privileged position to
efficiently describe the corner of Hilbert space. In fact, in 1D systems
at zero tempeature it is possible to
show: (i) that all gapped Hamiltonians with finite-range interactions fulfill the
area law \cite{Hastingsarea}; (ii) all states fulfilling the area law (even with logarithmic
corrections) can be efficiently approximated
by MPS with a number of parameters that only scales polynomially with $N$
\cite{verstraetecirac05} (see also \cite{osborne06}). This means
that for those systems at zero temperature we have found we were looking for and
explains why the celebrated DMRG
method introduced by White \cite{white92,white92b} gives extraordinary
approximations to the ground state of (finite) 1D spin chains.

A complementary approach to determine the interesting corner of Hilbert space has been
pursued by Hastings \cite{hastings06} (see also \cite{hastings06b}). Very
remarkably, he has been able to prove that, for any finite
temperature in any dimensions and Hamiltonian with finite-range interactions, PEPS efficiently describe
the thermal state in the sense that they approximate it arbitrarily close with a
polynomial number of parameters (Note that mixed state can be described in terms
of PEPS using the techniques of purification \cite{verstraeteripoll04,zwolak04}).
This shows that, for those problems, the corner
of Hilbert space has been identified. Note that the restriction of finite temperature
does not matter in many practical situations, since in any system we will always have that.
It is, nevertheless, interesting to investigate what happens at exactly
zero temperature. In that case, if one imposes a natural condition on the Hamiltonian
(related with how the density of states above the ground sate grow), Hastings has
also proven that PEPS provide an efficient description (note that for 1D systems
the area law already implies that \cite{verstraetecirac05}). In summary, MPS in 1D and
PEPS in higher dimensions provide us with accurate descriptions
of the states that appear in Nature under the conditions specified above
(most notably, short-range interactions).

\subsection{Tensor product states and renormalization group methods}

All the states above fall into a general class of states which can be called
TPS\footnote{Several authors have recently started calling this family of states
tensor network states (TNS). Here, for historical reasons, we will call them TPS.}
This class also contains states like Tree Tensor States
(TTS) \cite{FannesTTN} or
Multi-scale Renormalization Ansatz (MERA) \cite{vidal07a,vidal07b}.
It is characterized by the fact that the states are specified
in terms of a few tensors, of the order of $N$, each of them having a small rank
and low dimensions. Thus, they provide economic descriptions of certain quantum states.
During the last years a number of algorithms have been proposed to determine
those TPS for specific problems, giving rise to new numerical
methods suitable to describe certain many-body quantum systems. The purpose of the
present paper is to provide a pedagogical review of some of the most important TPS,
and to connect them to the successful real--space renormalization procedures that have
been used in Condensed Matter Physics for many years now. As we shall explain,
most of the TPS lie at the basis of such procedures. Making
the connection between those two problems (renormalization and tensor product
states) offers a new perspective for both of them, and explains the success
of some schemes from a different point of view. In particular, it relates the
structure of the states that appear in Nature under some conditions with the
ones that naturally appear in every renormalization procedure. We will mostly
concentrate on MPS and PEPS, but we will also
cover in part TTS and MERA, and briefly describe the rest. Note that, in the case
of PEPS, we have not been able to connect them to any known renormalization scheme.
Nevertheless, it may be interesting to find such a renormalization procedure since
it may give another insight in the field of Condensed Matter Physics. Furthermore, very
recently, several renormalization methods have been introduced in order to
determine expectation values of this and other kinds of TPS \cite{levin,levinwen,sandvik},
which is another evidence of the
strong bonds existing between these two fields.

This paper is organized as follows. Section II deals with MPS. We first introduce
them in terms of a renormalization procedure, and then explain two different methods
to carry out that procedure:
real-space renormalization \cite{wilson75} and density matrix renormalization group (DMRG)\cite{white92}.
We also present them from a quantum information perspective,
as the states that can be obtained by projecting pairs of entangled states
into lower dimensional subspaces. This provides a physical picture of the states,
which can be easily extended to higher dimensions and fermionic systems.
We also review how expectation
values can be efficiently determined, and introduce a graphical representation
of the state which strongly simplifies the notation and which will be used
later on instead of complex formulas. We show how MPS can be generated by a
sequential application of quantum gates \cite{schoen05} which, apart from providing us with
a specific recipe of how to engineer those states in practice, will allow
us to connect MPS and TTS in a simple way. Then we illustrate
how to approximate arbitrary states
by MPS. This provides the basis of several algorithms that have been
recently introduced to determine the ground state in different situations
\cite{verstraeteporras04,porras06,paredes05}, time evolution
\cite{cazalilla02,vidal04,whitefeiguin04,daley04,schollwoeck,verstraeteripoll04,sirker,Hastings09a,banuls09}, and thermal
states \cite{verstraeteripoll04,zwolak04,murg05}
of 1D spin chains. Finally, we introduce matrix product operators
\cite{verstraeteripoll04,zwolak04},
and show how they can be used to determine expectation values in a different way or to
describe thermal states via the purification procedure. Section III contains
a summary of a different renormalization method from which TTS naturally
emerge. As in the case of MPS, we show how one can determine expectation values,
and other quantities. In fact, it will become apparent that the algorithms
described in the previous section can be extended to TTS without
much effort. We also relate MPS and TTS showing that one can be expressed
in terms of the other one with a logarithmic effort. Then, in Section IV
we review yet another renormalization procedure \cite{Fisher,Fisherb} which
allows us to introduce the MERA \cite{vidal07a,vidal07b} from a different perspective.
As opposed to the TTS, MERA can be extended to higher
dimensions still fulfilling the area law, which makes them suitable
to study spin systems beyond chains. One way of generalizing MPS to higher dimensions is
through the PEPS, which are the subject of Section V. As opposed to the previous
sections, PEPS are not introduced in terms of a renormalization group procedure,
but following the intuition provided by the area law and the entanglement
present in the state. We finish the paper by reviewing some other extensions of
MPS to higher dimensions (for homogeneous and infinite systems).

Although we briefly mention how to build algorithms using TPS,
this is not the main purpose of this paper. For the reader
interested in the algorithms, rather than in the way they appear and
some of their properties, we recommend to have a look at Ref. \cite{reviewMurg}. For the reader
interested in the mathematical properties of MPS and PEPS, as well as the
development of a whole theory of many-body states based on that representation,
we refer to the paper \cite{Perezreview}.

In the following, we denote
the single spin Hilbert space by $\HH_1$, of dimension $d=2s+1$.
Spins interact with each other according to some Hamiltonian
$H=\sum_\lambda h_\lambda$, where $\lambda$ denotes some sets of
spins. As mentioned above, we will introduce different renormalization
procedures whose aim is to reduce the number of degrees of freedom
by putting together some spins and applying certain operators to
neighboring blocks of spins.

\section{Matrix Product States}

The standard Wilson Renormalization Group (RG) procedure as applied to
quantum impurity models can be viewed as follows \cite{wilson75,weichselbaum05}. We take
the first two spins, and consider a subspace of the corresponding
Hilbert space, $\HH_2\subset \HH_1\otimes \HH_1$ of dimension
$d_2\le d_1^2$. Now we add the next spin, and consider a subspace
of the Hilbert space of the three spins, $\HH_3\subset
\HH_2\otimes \HH_1$, of dimension $d_3\le d_2d_1$. We proceed in
the same way until we obtain $\HH_N\subset \HH_{N-1}\otimes
\HH_1$, and $d_N\le d_{N-1} d_1$. Now, we approximate the
Hamiltonian $H$ by $P_N H P_N$, where $P_N$ is the projector onto
$H_N$. If we can carry out this procedure, and we choose $d_N$
sufficiently small, we will be able to diagonalize this
Hamiltonian and find the eigenvalues and eigenvectors.

In order to be able to carry out this procedure in practice for a
large number of spins, we have to impose that all dimensions
$d_n\le D$, where $D$ is fixed (say, of the order of few
hundreds). We will call $D$ the {\em bond dimension}.
Otherwise we will run out of computer resources. Note
that, if we always take $d_n=d_{n-1}d_1$, we will end
up with the original Hilbert space of all the spins,
$\HH_1^{\otimes N}$. Although we will have made no approximation,
$d_n$ will grow exponentially with $N$, and thus the problem will
be untractable as soon as we have few tens of spins.

Before giving explicit recipes about how to properly choose the
subspaces at each step, and how to determine the final
Hamiltonian, let us show what will be the structure of the
eigenvectors of the approximate Hamiltonian $H_N$. Let us start
with $\HH_2$, and write an arbitrary orthonormal basis,
$\{|\beta\rangle_2\}_{\beta=1}^{d_2}$, in this subspace as
 \be
 \label{alpha2}
 |\beta\rangle_2 = \sum_{n_1,n_2=1}^{d_1} B^{n_1,n_2}_\beta
 |n_1\rangle_1\otimes|n_2\rangle_1.
 \ee
Here, $B^{n_1,n_2}_\alpha$ are the coefficients of the basis
vectors in terms of the original basis vectors $|n\rangle\in
\HH_1$. We can always express
 \be
 B^{n_1,n_2}_\beta=\sum_{\alpha=1}^{d_1} A[1]^{n_1}_{\alpha}
 A[2]^{n_2}_{\alpha,\beta}
 \ee
where $A[1]^{n_1}_{\alpha}=\delta_{n_1,\alpha}$ and
$A[2]^{n_2}_{\alpha,\beta}=B^{\alpha,n_2}_\beta$. The
orthonormality of the vectors (\ref{alpha2}), together with their
definitions, implies that
 \be
 \label{normal1}
 \sum_{n_1=1}^{d_1} A[1]^{n_1}_{\alpha}
 \bar A[1]^{n_1}_{\alpha'}=\delta_{\alpha',\alpha}, \quad
 \sum_{\alpha,n_2=1}^{d_1} A[2]^{n_2}_{\alpha,\beta}
 \bar A[2]^{n_2}_{\alpha,\beta'}=\delta_{\beta',\beta},
 \ee
where the bar denotes complex conjugate. We can now write any
orthonormal basis in $\HH_3$ in terms of linear combinations of
vectors $|\beta\rangle_2\otimes |n_3\rangle$, and proceed in the
same way iteratively. After $M$ steps, we have the relation
 \be
 \label{beta}
 |\beta\rangle_M = \sum_{\alpha=1}^{d_{M-1}}
 \sum_{n_M=1}^{d_1}
 A[M]^{n_M}_{\alpha,\beta}
 |\alpha\rangle_{M-1}\otimes|n_M\rangle_1.
 \ee
The orthonormality condition implies
 \be
 \label{normal2}
 \sum_{\alpha,n_M} A[M]^{n_M}_{\alpha,\beta}
 \bar A[M]^{n_M}_{\alpha,\beta'}=\delta_{\beta',\beta},
 \ee
Substituting recursively the definitions of $|\alpha\rangle_k$ in
(\ref{beta}), we can write
 \be
 \label{mps1}
 |\beta\rangle_N =
 \sum_{n_1,\ldots,n_N=1}^{d_{1}}
 (A^{n_1}_1 A^{n_2}_2\ldots A^{n_N}_N)_\beta
 \;|n_1,n_2,\ldots,n_N\rangle.
 \ee
Here, we have defined a set of matrices $A^n_M\in M_{d_{M-1},d_m}$
with components $(A^n_M)_{\alpha,\beta}:= A[M]^n_{\alpha,\beta}$
which, according to (\ref{normal2}), fulfill
 \be
 \label{normal3}
 \sum_{n=1}^{d_M} A_M^{n \dagger} A^n_M = \one
 \ee
i.e. the $A$'s are isometries. More specifically, when considering
the matrix $V$, of indices
 \be
 \label{isometry}
 V_{(n,\alpha)\beta}=A^n_{\alpha,\beta},
 \ee
where $(n \alpha)$ is taken as a single index, $V^\dagger V=\one$.
Note that, since $A^{n_1}_1$ are (row) vectors and the rest of the
$A$'s are matrices, the product $A^{n_1}_1 A^{n_2}_2\ldots
A^{n_N}_N$ is a vector, from which we take the $\beta$--th
component in (\ref{mps1}). The elements of the orthonormal basis
in $\HH_N$, and therefore all the vectors therein, can be
expressed in the form (\ref{mps1}): their coefficients in the
original basis $|n_1,\ldots,n_N\rangle$ can be written as product
of matrices. Vectors of that form are termed {\it matrix product
states} \cite{kluemper93} due to their structure. In particular, any state in
the coarse--grained subspace is a MPS which we will write as
 \be
 \label{mps2}
 |\Psi\rangle_N =
 \sum_{n_1,\ldots,n_N=1}^{d_{1}}
 A^{n_1}_1 A^{n_2}_2\ldots A^{n_N}_N
 \;|n_1,n_2,\ldots,n_N\rangle.
 \ee
Here, $A^{n_N}_N$ is a (column) vector fulfilling (\ref{normal3}),
i.e.
 \be
 \label{normal4}
 \sum_{n=1}^{d_{1}}\sum_{\alpha=1}^{d_{N-1}} |(A_N^{n})_{\alpha}|^2=1.
 \ee

Every MPS is invariant under the exchange of $A^n_M\to X_{M-1}
A^n_M X_{M}^{-1}$, where the $X$ are non--singular matrices, as it
can be checked by direct inspection of (\ref{mps2}). This gives us
the possibility of choosing a gauge, and thus impose conditions to
the matrices $A$ which simplify the further calculations, or which
give a physical meaning. In our case, we can consider
(\ref{normal3}) as a gauge condition which implements such a
choice and makes a direct connection between MPS and the
renormalization group method.

\subsection{Expectation values}

The whole renormalization procedure is completely determined by
the matrices $A^n_M\in M_{d_{M-1},d_M}$. Before giving different
prescriptions on how to determine them, we will show how to
calculate expectation values of different observables \cite{fannes92}. Let us
consider first $X=\sigma_1\otimes\sigma_2\otimes...\sigma_N$,
where the sigma's are operators acting on $\HH_1$. Using
(\ref{mps2}), we have
 \be
 \langle \Psi|X|\Psi\rangle =
 \sum_{n_k,m_k=1}^{d_{1}} \prod_k \langle m_k|\sigma_k|n_k\rangle
 C_{n_1,\ldots,n_M,m_1,\ldots,m_N}
 \ee
Here,
 \be
 C_{n_1,\ldots,n_M,m_1,\ldots,m_N}=A^{n_1}_1 \ldots A^{n_N} \bar A^{m_1}_1 \ldots \bar A^{m_N}_N
=(A^{n_1}_1\otimes\bar A^{m_1}_1)\ldots (A^{n_N}_N\otimes\bar
 A^{m_N}_N).
 \ee
Defining
 \be
 \label{EM}
 E_M[\sigma_M]:=\sum_{n,m=1}^{d_{1}} \langle m|\sigma_M|n\rangle
 A^{n}_M\otimes\bar A^{m}_M,
 \ee
we have
 \be
 \label{expval}
 \langle \Psi|X|\Psi\rangle = E_1[\sigma_1] E_2[\sigma_2] \ldots
 E_N[\sigma_N].
 \ee
Thus, the expectation value has a very simple expression: it is
itself a product of matrices \footnote{Note that $E_1$ and $E_N$
are a row and a column vector, respectively, and the rest
matrices}. Typically, one is interested in few--body correlation
functions, in which case most of the $\sigma$'s are the identity
operator. For those cases we will denote $E[\one]=:E$. In
particular, the normalization of $|\Psi\rangle$ can be written as
$\langle\Psi|\Psi\rangle= E_1 E_2 \ldots E_N=1$. This can be
readily checked by using (\ref{normal1},\ref{EM}) and writing the
vector $E_1$ as $(\Phi_1|$, where
 \be
 \label{PhiM}
 |\Phi_M):=\sum_{\alpha=1}^{d_M} |\alpha,\alpha),
 \ee
and the $|\alpha)$ are orthonormal vectors \footnote{Note that
they are in the space where the matrices $A$ act, and not in the
Hilbert spaces $\HH$ corresponding to the spins}. Besides that,
the orthonormality condition (\ref{normal2}) immediately implies
$(\Phi_{M-1}|E_M=(\Phi_M|$ and thus $E_1 E_2 \ldots
E_N=1$ [cf. (\ref{normal4})].

\subsection{Real--space renormalization group}

One of the simplest prescriptions to carry out the renormalization
procedure is to choose the orthonormal basis $|\beta\rangle_M$ as
we proceed \cite{wilson75,peschel}.
Since we are interested in the low energy sector of
our Hilbert space at each step, $\HH_M$, the natural choice is to
take the $d_M$ lowest energy states.

This construction becomes simpler if we have a Hamiltonian with
nearest-neighbor interactions only, i.e.
 \be
 \label{Ham1}
 H=\sum_{M=2}^{N} h_{M}
 \ee
where $h_n$ acts on spins $M-1$ and $M$, i.e.
 \be
 \label{hM}
 h_M=\sum_k \tilde h_M^{k,l} \sigma^{k}_{M-1}\otimes \sigma^{l}_M.
 \ee
We start out by choosing as $|\beta\rangle_2$ the $d_2$ eigenstates
of $h_2$ with the lowest energy, $e^{2}_\beta$. We can write
 \be
 h_2 \simeq \sum_{\beta=1}^{d_2} e^{2}_\beta |\beta\rangle\langle\beta|.
 \ee
In the $M$--th step, once we have determined the basis
$|\beta\rangle_{M-1}$ in the previous one ($M>2$), we write
 \be
 H_M=\sum_{n=2}^{M} h_{n}
 \ee
in the basis $|\beta\rangle_{M-1}\otimes|n_M\rangle$. This is very
simple, since we already have up to $h_{M-1}$ from the previous
step, and we thus just have to do that for $h_M$. Inspecting
(\ref{hM}) it is clear that, apart from the matrix elements of
$\sigma^l_M$ in the basis $|n\rangle$, we just have to determine
the matrix elements of any operator acting on the $(M-1)$ spin in
the basis $|\beta\rangle_{M-1}$. But this can be done right away,
since
 \be
 \langle\beta'|\sigma|\beta\rangle = \sum_{n=1}^{d_1}
 \sum_{\alpha=1}^{d_{M-2}} \langle
 m|\sigma|n\rangle
 (A^n_{M-1})_{\alpha,\beta}
 (\bar A^m_{M-1})_{\alpha,\beta'}.
 \ee
Once we finish this procedure we will end up with an orthonormal
basis in $\HH_N$, which should reproduce the low energy sector of
the Hamiltonian $H$.

\subsection{Density matrix renormalization group}

Starting from the previous discussion, we know that any real--space renormalization
procedure will give rise to a MPS. Thus, the best approximation to
the ground state of our Hamiltonian will be obtained using a
variational principle. That is, by minimizing the energy
$e=\langle\Psi|H|\Psi\rangle$ with respect to all MPS of the form
(\ref{mps2}). For that, we can just use the expression
(\ref{expval}) and write
 \be
 \label{HPsi}
 \langle\Psi|H|\Psi\rangle = \sum_{M=2}^N \sum_{k,l} \tilde h_M^{k,l}
 E_1 E_{2}\ldots E_{M-1}[\sigma^{k}]E_{M}[\sigma^{l}]E_{M+1}\ldots
 E_N.
 \ee
This formula explicitly shows the dependence of the energy on the
matrices $A^n_M$, and thus, in principle, can be used to determine
those matrices which minimize it. One possible strategy is to
minimize sequentially with respect to all possible $A$'s. That is,
we fix all $A^n_M$ for $M>2$, so that (\ref{HPsi}) is a function
of $A^n_2$ only\footnote{Note that $A^n_1$ is always fixed.}.
This functional dependence appears only in $E_2$ and
$E_2[\sigma^k]$. Actually, if we write (\ref{HPsi}) explicitly in
terms of $(A^n_2)_{\alpha,\beta}$, we see that such a dependence
is quadratic. We also have to impose the normalization condition
(\ref{normal3}), which in turn implies the normalization of
$|\Psi\rangle$, and which gives quadratic equations on those
coefficients as well. Thus, minimizing the energy at this point consists
of minimizing a quadratic polynomial with quadratic constraints.
Once we have $A_2^n$, we minimize with respect to $A_3^n$ by
fixing the rest of the $A$'s. We proceed in the same vein until we
reach $A_N^n$. At this point, we continue with $A_{N-1}^n$ and so
on. That is, we sweep the spins from left to right, to left, etc,
determining at each step the matrices associated to that
particular spin. At each step, the energy is smaller or equal than
the one considered in the previous step. This can be easily
understood since at that step we do not vary the matrices $A$ at
the site we are minimizing, we will obtain the previous energy.
Thus, minimizing with respect to the $A$'s cannot increase the
energy. Consequently, this procedure must converge to a minimum of
the energy \footnote{This can be a local or a global minimum.}

More specifically, each step consists of
minimizing
$\langle\Psi|H|\Psi\rangle/\langle\Psi|\Psi\rangle$.
Both numerator and denominator are quadratic functions of the
$A^n_m$ if we fix the rest. That is, if we write a long vector
$x_M$ containing all the coefficients of $A^1_M,
A^2_M,\ldots,A^{d_1}_M$, we can write
 \be
 e= \frac{x_M^\dagger H_M x_M}{x_M^\dagger N_M x_M},
 \ee
where $H_M$ and $N_M$ can be determined using (\ref{HPsi}). It
follows from the hermiticity of $H$ that $H_M$ is hermitian.
Furthermore, $N_M$ is positive semidefinite (as the norm of a
vector must be non--negative). Thus, $e$ is real and its minimum
can be determined through the generalized eigenvalue equation
 \be
 \label{genereig}
 H_M x_M = \lambda N_M x_M,
 \ee
where one has to choose the minimum $\lambda$ fulfilling that. From
this equation, we also determine $x_M$ and thus $A^n_M$. The
only drawback is that the matrix $N_M$ may be ill--conditioned,
something which may pose some problems when solving Eq.
(\ref{genereig}).

There is a way to circumvent this misfortune and, at the same
time, simplify the algorithm further. As mentioned above, we have
that $(\Phi_{M-1}|E_M=(\Phi_M|$, a fact that can be used to
simplify Eq. (\ref{HPsi}). Actually, we could have started our
renormalization procedure from the $N$--th spin in decreasing
order. In that case, instead of (\ref{normal2},\ref{normal3}) we
would have had $E_N=|\Phi_1)$,
 \be
 \label{normal5}
 \sum_{n=1}^{d_M} A^n_M A_M^{n \dagger} = \one,
 \ee
and thus $E_M|\Phi_{M+1})=|\Phi_M)$. This suggests a mixed
strategy, where we renormalize in increasing order up to the spin
$M-1$, and in decreasing one up to $M+1$, whenever we are
minimizing the energy in site $M$. In this case we have
 \bea
 \langle
 \Psi|\Psi\rangle&=& E_1 \ldots E_{M-1} E_M E_{M+1}\ldots E_N=
 (\Phi|E_M|\Phi)\nonumber\\
 &=& \sum_{\alpha,\beta=1}^{d_{M-1},d_{M}} \sum_{n=1}^{d_1}
 |(A^n_M)_{\alpha,\beta}|^2.
 \eea
That is, the matrix $N_M$ is simply the identity matrix, and
therefore Eq. (\ref{genereig}) becomes a standard eigenvalue
equation. Once we have determined $A^n_M$ by solving it, and if we
are going to minimize next the matrices at site $M+1$, we write
$A^n_M=U^n_M X$, where $U^n_M$ is an isometry [i.e. fulfills
(\ref{normal3}), cf. (\ref{isometry})]. For instance, we
define $V$ according to (\ref{isometry}), and use the singular
value decomposition to write $V=U D W$, where $U$ and $W$ are
isometries and $D\ge 0$ is diagonal, i.e. $X=DW$. We can now discard
the $X$ and keep $U^n_M$ for site $M$, since we can always take
$A^n_{M+1}\to X A^n_{M+1}$, so that the state $|\Psi\rangle$
remains the same. Given that we are going to optimize now anyway
with respect to $A^n_{M+1}$, we can ignore the multiplication by
$X$. In this way, we make sure that $(\Phi_{M-1}|E_M=(\Phi_M|$,
which is consistent with our procedure. In case we are going to
minimize next the matrices at site $M-1$, we just have to
decompose $A^n_M=X U^n_M$, where now $U^n_M$ fulfills
(\ref{normal5}). Apart from that, it is numerically convenient to
store the values of $E_k[\sigma]E_{k+1}[\sigma']E_{k+2} \ldots
E_{M-1}$ and $E_{M-1}[\sigma]$, for the $\sigma$'s that appear in
the Hamiltonian, and update them when we have determined $A^n_M$.
Besides, one should also store those of the form $E_{M+1}\ldots
E_{k-1}E_{k}[\sigma]E_{k+1}[\sigma']$ as well as
$E_{M+1}[\sigma]$, since they will be needed in future
optimizations.

The algorithm carried out in this way is (up to minor modifications) the
celebrated {\it density matrix renormalization group} algorithm
introduced by S. White in 1991 \cite{white92,white92b}. The only minor differences are:
(i) he optimized two sites at the same time, say $M$ and $M+1$ and
from there he determined $A^n_M$ by using a singular value
decomposition; (ii) he proposed a method to determine the initial
configuration of the $A$'s by growing the number of spins until
the desired value $N$ is reached. Furthermore, he derived
his algorithm from a very intuitive method in which he described the
effect of the rest of the spins on any given one in the normalization
in an efficient
way. In any case, the present discussion highlights the variational
character of (the finite version of) DMRG \cite{verstraeteporras04}.

\subsection{Matrix Product States and Projected Entangled-Pair
States}

The MPS can also be introduced from a different perspective, which
highlights their entanglement content and is amenable to several
generalizations. The idea is to extend the AKLT construction \cite{AKLT1},
where one substitutes the original spins by couples of auxiliary
systems in a prescribed state, and then projects their state back
to the spin Hilbert space \cite{verstraetecirac04,verstraetecirac04b}.

\begin{figure*}
\centering
\includegraphics[scale=0.7]{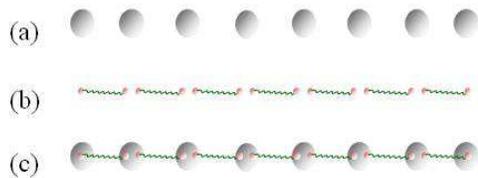} \\
\caption{\label{fig1}%
  Construction of a MPS in terms of entangled auxiliary
  particles. (a) Original spin system; (b) We replace each
  spin by two auxiliary particles (except at the ends of the chain),
  which are in a maximally entangled state with their neighbors; (c)
  The final state is obtained after mapping the state of each pair of
  auxiliary particles locally onto the original spins.}
\end{figure*}

For the definition and construction of the state, we imagine that
each site of the spin chain, $M$, we have two ancillas $l_M$,
$r_M$, with corresponding Hilbert spaces of dimensions $d_{M-1}$
and $d_M$, respectively (see Fig.
\ref{fig1}(a)). For $M=1$ ($M=N$) we have only one ancilla
$r_1$ ($l_{N}$). The state of the ancillas is fixed: they are
maximally entangled to their nearest neighbor (see Fig.
\ref{fig1}(b)). That is, $r_{M-1}$ and $l_{M}$ are in a state
$|\Phi_{M-1})$\footnote{We have used $|...)$ to denote states on
the Hilbert space of the ancillas. As it will be clear, this is a
natural choice in view of our definition (\ref{PhiM}).} Now, in
order to recover the MPS, we map the state of the ancillas onto
the one of the spins at each site (Fig. \ref{fig1}(c)). That is, we write
 \be
 \label{P1}
 |\Psi\rangle = P_1\otimes\ \ldots P_N
 |\Phi_1)\otimes\ldots |\Phi_{N-1})
 \ee
where $P_M:\HH_{M-1}\otimes \HH_M \to \HH_1$. We write each of
those maps in the bases we have chosen for each Hilbert spaces
 \be
 P_M = \sum_{n=1}^{d_1} \sum_{\alpha=1}^{d_{M-1}}
 \sum_{\beta=1}^{d_{M}} (A^n_M)_{\alpha,\beta}
 |n\rangle(\alpha,\beta|.
 \ee
It is a simple exercise to show that, indeed, the state defined in
(\ref{P1}) coincides with that of (\ref{mps2}). Thus, MPS can be
obtained by projecting entangled-pairs of ancillary particles onto
the spaces of the spins, thus the name 1--dimensinal PEPS.

This construction allows us to derive a variety of the properties of MPS right
away. First of all, if we consider the reduced density operator,
$\rho_{1,2,\ldots,M}$, of the first $M$ spins, its rank (number of
non-zero eigenvalues) is bounded by $d_M$. The reason is that the
original state of the ancillas obviously fulfills that condition,
and that the rank cannot increase by applying any operator, in
particular $P_1\ldots\ldots P_M$. This implies that the
von--Neumann entropy $S(\rho_{1,2,\ldots,M})\le \log_2 d_{M}$
\footnote{We take $S(\rho)=-{\rm tr}[\rho\log_2\rho]$}, i.e. the
entropy of a block of contiguous spins is bounded by the maximum
value of $d_M$. The fact that the entropy of a block of spins is
bounded (for infinite chains) is sometimes referred as the area
law, as mentioned in the introduction.

Apart from that, we can also prove that any spin state can be
written as a MPS. For that, consider that $l_M$ ($r_M$) is, in
turn, composed of $N-M+1$ ($N-M$) spins, as sketched in Fig.
{\ref{fig1}. Now, for $M>1$ the map
 \be
 P_M=(\Phi_2|^{\otimes N-M}\otimes \sum_{n=1}^{d_1} |n\rangle(n|
 \ee
teleports \cite{teleport} the state of the first $N-M$ spins of $l_M$ to
$r_{M+1}$, while leaving the last one as the physical spin. By
choosing
 \be
 P_1 = \sum |n\rangle(\Psi_n|\otimes (n|
 \ee
where $|\Psi_n\rangle = _1\langle n|\Psi\rangle$, we have the
desired result.

\subsection{Translationally invariant systems}

We mention now the possibility of dealing with
systems with periodic boundary conditions corresponding, for instance, to
translationally invariant Hamiltonians. A MPS which automatically
fulfills this condition is the one in which the matrices $A$ at
different sites are the same. In that case, we can write
 \be
 \label{mpsperiodic}
 |\Psi\rangle_N =
 \sum_{n_1,\ldots,n_N=1}^{d_{1}}
 {\rm tr}[A^{n_1} A^{n_2}\ldots A^{n_N}]
 \;|n_1,n_2,\ldots,n_N\rangle.
 \ee
This class of states, in the limit $N\to\infty$, appeared even before
the name MPS was coined, and were called {\em finitely correlated
states}\cite{fannes92}. They were introduced when extending the
1D version of the AKLT model \cite{AKLT1}, whose ground state is the most prominent
example of a state in that clase. They were also viewed as
a systematic way of building translationally
invariant states\cite{fannes92}. States with different matrices were not
considered at that time since the emphasis was given to infinite translationally
invariant spin chains. Note also, that any translationally invariant state
may be written in this form \cite{perezgarcia06}.

The matrices $E_M=:E$ now coincide and $E$ is called transfer
matrix, given the analogy of the formulas with those of classical
statistical mechanics. Its eigenvalues reflect the correlation
length and other properties of the system. This can be understood
since if we consider the two--spin (connected) correlation functions
at distance $L$, the matrix $E^L$ will appear in the
calculation. In the limit $L\gg 1$ only the largest eigenvalues of
$E$ will give a contribution to the correlation function, which
will thus decay exponentially. Thus, the name finitely correlated
states used for states of the form (\ref{mpsperiodic}) in the
limit $N\to \infty$ \cite{fannes92}.

Let us show that, without loss of generality, we can impose the
Gauge condition ({\ref{normal3}). For that, we denote by $X$ the
operator corresponding to the largest eigenvalue $\lambda$ (in absolute
value)
of the following eigenvalue equation
 \be
 \sum_{n} A^{\dagger n} X A^n=\lambda X.
 \ee
One can can always choose $\lambda=1$ by re-scaling and $X=X^\dagger >0$. Then,
the matrices $\tilde A^n:=X^{1/2} A^n X^{-1/2}$
are well defined, correspond to the same state as that of $A^n$,
and fulfill the Gauge condition.

We consider now the renormalization procedure in this particular
case. Instead of performing that step by step, as explained above,
we can aim at minimizing the energy directly within the MPS of the
form (\ref{mpsperiodic}). For that, we just have to specialize
(\ref{mpsperiodic}), and consider a single term in the Hamiltonian
(\ref{hM}) given the translational symmetry, i.e.
 \be
 \label{HPsi2}
 \frac{\langle\Psi|H|\Psi\rangle}{\langle\Psi|\Psi\rangle}
 = N \sum_{k,l} \tilde h_M^{k,l}
 \frac{{\rm tr}(E[\sigma^{k}]E[\sigma^{l}]E^{N-2})}{{\rm tr}(E^{N})}
 \ee
Note that, as opposed to the previous case (where we had open
boundary conditions), the Gauge condition does not guarantee that
the state $|\Psi\rangle$ is normalized. The energy so defined is a
function of the $A$'s, so that we can aim at minimizing this
expression directly. The latter simplifies in the limit
$N\to\infty$ if the maximum eigenvalue of $E$ is not degenerate,
 \be
 \label{HPsi3}
 \lim_{N\to\infty} \frac{\langle\Psi|H|\Psi\rangle}{N\langle\Psi|\Psi\rangle}
 = \frac{1}{\lambda}\sum_{k,l} \tilde h_M^{k,l} \langle L|
 E[\sigma^{k}]E[\sigma^{l}]|R\rangle.
 \ee
Here
$|L\rangle$ and $|R\rangle$ are the right and left
eigenvectors corresponding to the maximum eigenvalue $\lambda$
 \be
 E|R\rangle =\lambda|R\rangle,\quad \langle L|E=\lambda\langle L|.
 \ee
Here we have not imposed the Gauge conditions, in which case we
would have $\lambda=1$. The minimization of (\ref{HPsi3}) with
respect to $A^n$ can be now performed directly using, eg,
using conjugate gradient methods.

The variational method exposed above was first proposed by
Rommer and Ostlund in the context of DMRG \cite{oestlund95,rommer97}
(see also \cite{dukelsky98}). They realized that, in the infinite version
\cite{white92}, if $N\to \infty$, the fixed point of the DMRG procedure
will correspond to a MPS with all matrices equal (ie a finitely correlated
state). Then they suggested to take those states as variational states
and minimize the energy using standard methods. Alternatively,
one may perform the minimization
by using evolution in imaginary time applied to an infinite system \cite{vidal07}.
It is not strictly necessary to take all
the matrices $A$ identically (i.e. $\ldots AAAA\ldots$) \cite{reviewMurg}, but it may be
more convenient to alternate two kinds of matrices (i.e. $\ldots
ABAB\ldots$)\cite{vidal07}. Note that this case would be included if we group pairs
of neighboring spins and take identical matrices in each group ($CCCC\ldots$),
since this covers the previous case if we take $C=AB$. In practice, however,
the latter approach may be less efficient numerically since one has
to deal with larger spins.

\subsection{Graphical representation}

When using many-particle quantum states, we reach very soon a cumbersome level
of notation. This is not an exception when we utilize the language of
MPS, since we typically have products of many matrices, which depend on
another index that corresponds to each individual spin. As soon as we express
expectation values of observables, the notation gets very involved. There is
a simple way of conveying the same information by using a graphical
representation of MPS, which we will introduce here and that will be used
in the following sections. For that, let us consider the $A$'s describing
the MPS as a rank three tensor ($A^n_{\alpha,\beta}$ of indices $n,\alpha,\beta$),
as shown in Fig. \ref{fig2}(a). These tensors are contracted along the
indices $\alpha,\beta$ to form the matrix product state (Fig. \ref{fig2}(b)).
More precisely, the $\langle n_1,\ldots,n_N|\Psi\rangle$ is obtained after
this contraction, where the indices $n$ are still open. We can thus represent the
MPS as the tensor of Fig. \ref{fig2}(c). If the first and the last objects
are also rank three tensors, we will have the representation of Fig. \ref{fig2}(d),
which in turn describes, eg, a translationally invariant state.
Any local observable, $\sigma$, can be represented as a tensor itself, if we write
it in the spin basis (Fig. \ref{fig2}(e)).

\begin{figure*}
\centering
\includegraphics[scale=0.7]{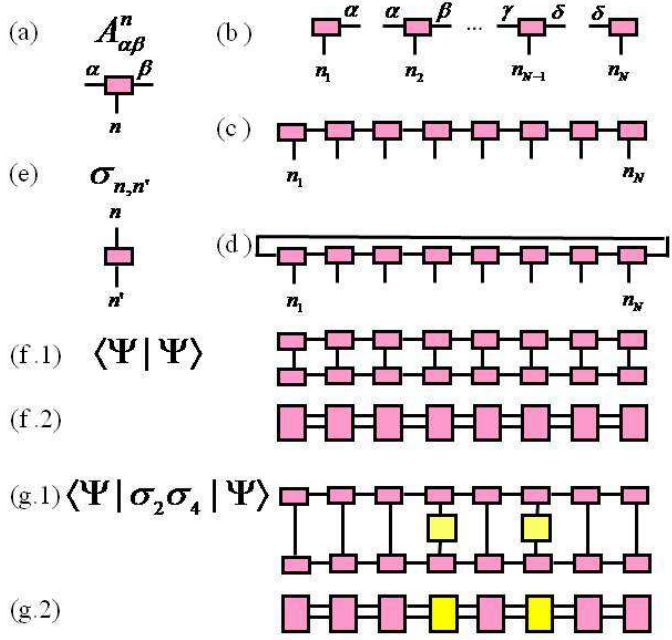} \\
\caption{\label{fig2}%
  Graphical representation of an MPS in terms of contracted tensors (tensor
  network). (a) The set of matrices $A^n$ are represented in terms of a rank--3
  tensor where the index $n$ is pointing vertically; (b) We consider the set of
  tensors corresponding to each spins and (c) contract them according to the
  horizontal indices; (d) the same can be done with periodic boundary conditions
  by adding an extra bond on the end spins; (e) Tensor representation of
  an operator acting on a spin; (f.1) In order to calculate $\langle \Psi|\Psi\rangle$
  we contract the tensor corresponding to $\Psi$ with that of $\bar\Psi$, giving
  rise to (f.2) a row of tensors which are contracted to give a number. The
  tensors can be viewed as matrices (one double-index to the left and another to
  the right). (g.1) and (g.2) the same but with an expectation value.}
\end{figure*}

The norm of the state can be obtained by tracing the tensor with respect to the
spin indices. This is represented in Fig. \ref{fig2}(f.1), where the upper part represents
$\langle n_1,\ldots,n_N|\Psi\rangle$ and the lower the complex conjugate, and the
indices $n$ are contracted. By considering each pair of tensors $A$ and $\bar A$ on top
of each other, we can build the matrix $E[\one]$ defined in (\ref{EM}), and thus
represent the norm as the contraction of those matrices (Fig. \ref{fig2}(f.2), compare
(\ref{expval})). In the
same way, we can represent expectation values of product of local observables
(Figs. \ref{fig2}(g.1) and (g.2)).

\subsection{Sequential generation of Matrix Product States}

We have seen so far that the family of MPS corresponds to those that appear in real space
renormalization schemes. Here we will show that they also coincide with the states that can be
sequentially generated\cite{schoen05}. For that, let us assume first that we have an auxiliary
system, i.e an ancilla (which,
in practice, could be a $D$--level atom) with Hilbert
space $H_a$ of dimension $D$, initially prepared in state $|1\rangle$, and also all the
spins in the chain in state $|1\rangle$. Now we consider a unitary operation between
the ancilla and the first spin, then between the ancilla and the second on, and so on, until the ancilla interacts
with last spin (see Fig. \ref{fig3}(a)). Let us denote by $U^{(A,M)}$ the unitary operation
between the ancilla and the $n$--th spin. If we now denote by $|\alpha)$ an orthonormal
basis in the ancilla Hilbert space, and write
 \be
 \label{UAM}
 U^{(A,M)} |0\rangle_1 = \sum_{\alpha,\beta=1}^D\sum_{n=1} A[M]^n_{\alpha,\beta} |n\rangle_1
 |\alpha)(\beta|,
 \ee
then we end up with the state
 \be
 \label{mpsancilla}
 |\Psi\rangle_N =
 \sum_{n_1,\ldots,n_N=1}^{d_{1}}
 (A^{n_1}_1 A^{n_2}_2\ldots A^{n_N}_N)_\beta
 \;|n_1,n_2,\ldots,n_N\rangle|\beta).
 \ee
In Fig. \ref{fig3}(b) we have used the representation introduced above to describe
this process and to arrive at the above formula.

\begin{figure*}
\centering
\includegraphics[scale=0.7]{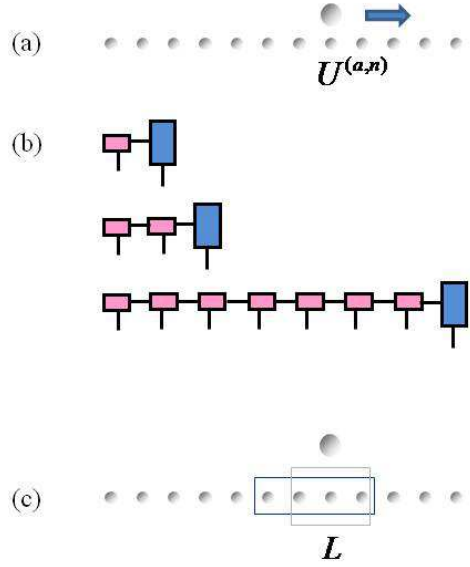} \\
\caption{\label{fig3}%
  Sequential generation of MPS. (a) Using an ancilla with Hilbert space of
  $D$ dimensions, we act sequentially on the first, second, etc spins with
  unitary operators; (b) This process can be understood with the graphical
  language introduced before. After each interaction, the spins get entangled
  in a MPS with the ancilla itself. (c) We can replace the ancilla by the
  $log D$ spin which are to the right of the spin we are acting on.}
\end{figure*}

From Eq.(\ref{mpsancilla}) it also immediately follows that any MPS can be created
sequentially using an ancilla, by simply choosing the $U^{(A,M)}$ according to
(\ref{UAM}), and the last one in such a way that $A^{n_N}_N$ is a vector (ie it only
has one component $\beta=1$). Besides, we can substitute the ancillary particle at each
step by $L\ge \log_{d_1}D$ spins which lie on the right of the particle we are acting
on, since they span a Hilbert space which has dimension at least $D$. Once we apply
$U^{A1}$ to the first $L+1$ spins, we can swap the state of the spins $2,\ldots,L+1$ to
$3,\ldots,L+2$ by using a unitary operation acting on spins $2,\ldots,L+2$. Now we can
use the spins $3,\ldots,L+2$ as ancilla, and apply $U^{A2}$ to the second spin plus
those. Proceeding in this way, we see that we can also prepare any MPS by a sequence of unitary
operations from left to right, each of them acting on at most $L+1$ spins (see Fig. \ref{fig3}(c)).

\subsection{Approximating states with Matrix Product States}

Given a state, $|\Phi\rangle$, expressed in a given basis, how do we find its MPS representation?
This is very simple in theory, since we just have to follow the procedure
of the renormalization, but keeping all the states that are necessary. We can do that
as follows. We first consider the reduced density operator of the first spin,
$\rho_1={\rm tr}(|\Phi\rangle\langle\Phi|)$ and a basis, say $|\beta\rangle_1$, where it is
diagonal. Then we consider the first two spins, and do the same with the reduced
density operator $\rho_2$, determining a basis $|\beta\rangle_2$. Obviously, we can
write (\ref{alpha2}). We continue in the same way, so that at the end we can
write the state $|\Phi\rangle$ as in (\ref{mps1}), and thus as a MPS.

This procedure, in general, will not work with a large
number of particles since we will have to diagonalize matrices ($\rho_M$)
of dimensions that grow exponentially with the number of spins, and thus the
matrices $A$ will also be too large to be handled (Note that if
the rank of all operators $\rho_M$ is smaller than some fixed number, say $D$,
the procedure will express the state as a MPS with maximal bond dimension $D$.
In other words, any state can be written as a MPS of bond dimension equal to
the maximal rank of the reduced density operators $\rho_M$). One way to circumvent
this problem is to look for a good approximation to the state $|\Phi\rangle$ in
terms of a MPS, $|\Psi\rangle$. One can do that using the same idea as the renormalization procedures
presented above:

\begin{itemize}

\item {\em Real-space-like approximation:}  \cite{vidal03} Here, every time we
diagonalize
$\rho_M$, we only pick $D$ states $|\alpha\rangle_M$, those eigenvectors
corresponding to the $D$ largest eigenvalues  of $\rho_M$. In this way,
as in the real-space renormalization procedure, we try at each step to
be as close as possible to the state $|\Phi\rangle$.

\item {\em DMRG-like approximation:} \cite{verstraeteripoll04} Instead of trying to optimize locally,
at each step, the subspaces we select in order to represent $\Phi$, we can
do something better. The inspiration comes from DMRG, where one does not
perform an optimization locally, but more globally. We can do the same thing here,
ie to obtain the $A$'s that approximate
the state $|\Phi\rangle$ variationally, so that they provide the best possible
approximation. In practice,
this means that we maximize $|\langle \Psi|P|\Psi\rangle|$, where
$P=|\Phi\rangle\langle\Phi|$. This minimization is, in turn, similar to the
minimization of the energy of the ground state of $H$, cf. Eq. (\ref{HPsi}). Thus,
we can follow the same procedure, namely sequentially minimize with respect to
each of the $A$'s fixing the rest.

\end{itemize}

The procedures exposed above will be still hard to implement in practice, due to the fact
hat we still have to deal with too many parameters (those describing $|\Phi\rangle$).
However, if that state is initially written in a MPS form (of matrices with a large
but fixed bond dimension, $D$) or in a superposition thereof
with few terms, then we will be able to do that in practice, since all the
operations can be made efficiently. For example, the term $|\langle \Psi|P|\Psi\rangle|$
will be a polynomial of second degree in the coefficients of each particular
 $A[M]$ that can be easily determined.

The above procedure can be used to simplify and compress a MPS description. For
instance, imagine we have a MPS with matrices $A$ of dimension $D$. The goal is to
find another MPS which is very close to that one, of matrices $B$ with a smaller
dimension, $D'$. In practice, this technique can be used for many purposes. For example,
to approximate the evolution of a MPS under the action of a Hamiltonian of the form
(\ref{Ham1}), or some quantum gates. The idea in this case is to apply the evolution
operator for a short time, such that we can determine the evolution after this step
(using, eg, perturbation theory or neglecting terms that do not commute in the Hamiltonian),
and then approximate that state with a MPS of a fixed dimension. The real-space
approximation explained above gives rise to the celebrated time evolving block decimation (TEBD)
algorithm
\cite{vidal04,whitefeiguin04,daley04,schollwoeck} whereas the optimal DMRG-like one was
presented in \cite{verstraeteripoll04}.

\subsection{Matrix Product Operators}

In the same way one defines MPS, one can define operators which can be written
in terms of products of matrices, the so-called matrix product operators (MPO)
\cite{verstraeteripoll04}. They are of the form
 \be
 \label{mpo}
 X =
 \sum_{n_1,\ldots,n_N=1}^{d_{1}}
 {\rm tr}(B^{n_1}_1 B^{n_2}_2\ldots B^{n_N}_N)
 \;O_{n_1}\otimes O_{n_2}\ldots \otimes O_{n_N}.
 \ee
Here, $\{O_n\}_{n=1}^{d^2_1}$ forms a basis in the space of operators acting on $\HH_1$.
For instance, we may take $O_{nn'} =|n\rangle\langle n'|$.

A class of operators that can be easily written as MPO are Hamiltonians with
short-range interactions. The idea is to realize that the space of operators
acting on each spin is itself a Hilbert space (now of dimension $d^2_1$), so
that all the properties of MPS directly apply to MPO, and this allows us to write
the Hamiltonian as a simple MPO. In order to do that explicitly, let us write
$|k\rangle=\sigma^k$ with $|1\rangle=\one$, so that $h_M$ defined in (\ref{hM})
becomes
 \be
 \label{hMp}
 h_{M}= |1\rangle \otimes ...|1\rangle \otimes \left[\sum_k \tilde h_M^{k,l}
 |k\rangle_{M-1}\otimes |l\rangle_M\right] ... \otimes |1\rangle.
 \ee
 From this expression it becomes apparent that $h_M$ can be written as a MPS
with bond dimension, $D\le d_1^2$: when considering $H$, we can take it as a sum
of three terms, those corresponding to $M<M_0$, $M>M_0$, and $M=M_0$. When writing
the analog of the reduced density operator, $\rho_{M_0}$ for $H$ we see that both the
first term and the second one will just give a contribution of one to its rank,
whereas the third one gives at most $d_1^2-2(d_1-1)$
(Note that we can
include the terms with $k=1$ or $l=1$ in $h_{M_0\mp 1}$, respectively).
Thus, $H$ can be written as a MPO with bond dimension $D\le d^2_1-2d_1+3$.
If we write the Hamiltonian (or any other operator) in this form, one can determine expectation values
with MPS in a very efficient way (compare (\ref{expval})):
 \be
 \langle\Psi|X|\Psi\rangle=\tilde E_1 \tilde E_2 \ldots
 \tilde E_N,
 \ee
where
 \be
 \tilde E_M = \sum_{n,m,k} A^n_M\otimes B^k_M\otimes \bar A^m_M \langle m|O_k|n\rangle.
 \ee
This provides an alternative way of determining expectation values of the
Hamiltonian, and thus to carry out DMRG calculations.

Another class of operators for which it is useful the MPO description is the one
of density operators, $\rho$, describing the full spin chain. Those are self-adjoint and
positive semidefinite, something
which is not easy to express in terms of the matrices $B$. However, for those we can
use the idea of purification: namely, we can always extend our spin chain with
another auxiliary one, with the same number of spins, in such a way that the
state of both chains, $|\Psi\rangle$, is pure but when we trace the auxiliary one we obtain the
original density operator, $\rho$ \cite{verstraeteripoll04}.
We write $|\Psi\rangle$ as an MPS
 \be
 |\Psi\rangle =
 \sum_{n_1,\ldots,n_N=1}^{d_{1}}
 A^{n_1m_1}_1 A^{n_2m_2}_2\ldots A^{n_Nm_N}_N
 \;|n_1,n_2,\ldots,n_N\rangle|m_1,m_2,\ldots,m_N\rangle.
 \ee
Thus, we obtain $\rho$ in the form of a MPO (\ref{mpo}) with
$O_{nn'} =|n\rangle\langle n'|$ and
 \be
 B^{nn'}_M = \sum_{m=1}^{d_1} A^{n,m}_M\otimes \bar A^{n'm}_M.
 \ee
Now, we can use the methods described in previous sections to determine
the time evolution of a density operator (either directly using the MPO description,
or using the purified state). In particular, we can describe the thermal
equilibrium state as
 \be
 e^{-\beta H} = e^{-\beta H/2}\one e^{-\beta H/2},
 \ee
where $\one$ can be trivially expressed in terms of a purification. Thus, by performing
the time evolution (in imaginary time) starting from the purification of $\one$ up
to a time $t=i\beta/2$, we obtain the desired MPO.

Other examples of MPO are transfer matrices in classical systems, as well
as the monodromy matrices as appearing in the algebraic Bethe Ansatz \cite{frank}.

\section{Tree Tensor States}

Another way of carrying out the renormalization procedure in one
dimension is to follow Kadanoff's original idea \cite{kadanoff}. Let us assume,
for simplicity, that $N$ is a power of 2. We split our $N$ spins
into $N/2$ neighboring pairs. For each pair, we consider a
subspace $\HHk_2\subset \HH_1\otimes \HH_1$ of dimension $d_2^{\rm
k}$. The resulting systems are paired again into $N/4$ neighboring
couples, where we take $\HHk_3\subset \HHk_2\otimes \HHk_2$ of
dimension $d_3^{\rm k}$, and continue in the same vein until we
end up with a single system. The comparison of this way of performing
the renormalization and the one given in the previous section is
represented in Fig. \ref{fig4}.

\begin{figure*}
\centering
\includegraphics[scale=0.7]{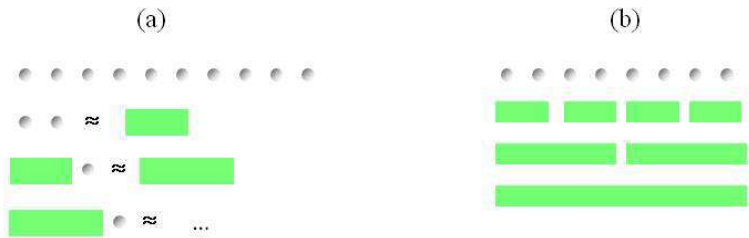} \\
\caption{\label{fig4}%
 Comparison of the two renormalization procedures. (a) At each
 step, we add a new spin (ball) to the previous system (square)
 obtaining a new Hilbert space, which we truncate to obtain the
 one of a smaller dimension (new square). (b) At each step, we take
 two neighboring systems (squares) and truncate the Hilbert space
 to obtain the new one of the new systems.}
\end{figure*}

As before, we can follow which kind of states are supported in the
final subspace $\HHk_n$. At the second step, an orthonormal basis in $\HHk_2$ corresponding to
particles $2M-1$ and $2M$ ($M=1,\ldots,N/2$) is
 \be
 |n\rangle_M = \sum_{n_1,n_2=1}^{d_1} (T_M^{2})^{n}_{n_1,n_2} |n_1\rangle_{2M-1}
 \otimes |n_2\rangle_{2M},
 \ee
where $n=1,\ldots,d_2^{\rm k}$. In the $i$--th step, we can use the same formula
to express the basis in $\HHk_i$ in terms of $\HHk_{i-1}$ just by replacing
$T^{(2)}$ by $T^{(i)}$. The orthonormality of the basis gives us the condition
 \be
 \label{Torth}
 \sum_{n_1,n_2=1}^{d_{i-1}} \bar (T_M^{(i)})^{n'}_{n_1,n_2} \bar (T_M^{(i)})^n_{n_1,n_2}=\delta_{n',n},
 \ee
ie, $T$ must be an isometry.
The final state can be easily written in terms of the $T^{(i)}$, but we will not
do that here. Instead, we will use a graphical representation as we did in the case of MPS.

Let us start out representing the tensors $T$ as in Fig. \ref{fig5}(a). The orthonormality
condition (\ref{Torth}) can be thus graphically illustrated as in Fig. \ref{fig5}(b). That is,
contracting the indices $n,m$ of the tensor with its complex conjugate gives a delta
function (here represented by a line). Any state $|\Psi\rangle$ obtained with this renormalization
scheme will have the structure of Fig. \ref{fig5}(c). That is, it will consist of
different isometries $T$, characterizing the truncation of the Hilbert space of pairs
of subsystems, which are contracted according to the diagram
\cite{FannesTTN}. Note that we could have
joined more than two spins in the first step, or in successive steps, in which case we
would have obtained a similar diagram but in which the tensors would have more indices.
One calls this class of states Tree Tensor States (TTS) since the diagram resembles a tree.

\begin{figure*}
\centering
\includegraphics[scale=0.7]{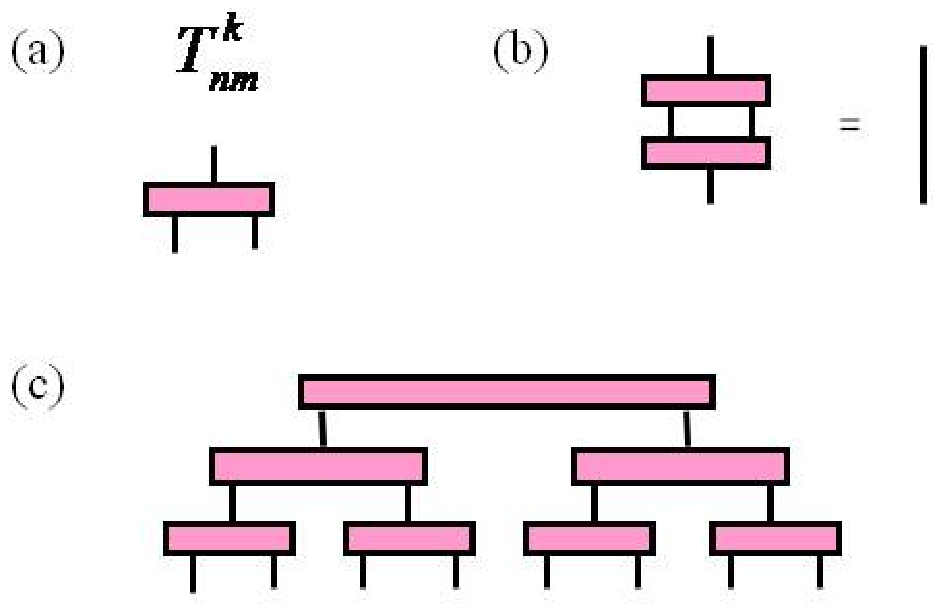} \\
\caption{\label{fig5}%
 Tensor network representation of the TPS. (a) Each tensor $T$ is represented by
 a square with three indices. (b) the fact that $T$ is an isometry can
 be represented as a line when we contract $T$ and $\bar T$. (c) A TTS. }
\end{figure*}

By looking at the tree structure of a state (Fig. \ref{fig5}(c)) it is very easy to notice
that the states may violate the area law. In fact, if we look at a block of contiguous spins,
we can deform the diagram and see how many links connect that block with the rest. Depending on
where we take the block, the number may vary. In the figure, for instance, if we take a block with
spins $2M-1$ and $2M$, then they will be connected by a single bond to the rest of the spins.
If we take instead the spins $2$ and $3$, they are connected by two bonds to the rest.
It is easy to realize, as in the case of MPS, that the entropy of the block is bounded by
the sum of the logarithms of the dimension of the bonds that connect the block with the rest.
For some block, this entropy is bounded by a constant, whereas for some other ones
it is bounded by $c \log L$, where $L$ is the number of spins in the block. Thus, a TTS
violates the area law, although only mildly. In fact, given that critical systems
typically have a logarithmic correction to the area law \cite{calabrese}, it is natural to try to describe
them with TTS, and thus with the renormalization group procedure exposed here,
as it is usually done.

\subsection{Expectation values}

Expectation values of observables in a TTS can be easily evaluated, as it
was the case for MPS. Let us take, for example, an observable acting on a give
site, ie, $\langle \Psi|\sigma|\Psi\rangle$. We can write this expectation value
as a contraction of two TTS, $\Psi$ and $\bar\Psi$, in which the operator
$\sigma$ is sandwiched in between (Fig. \ref{fig6}(a)). Using the fact that
all $T$'s are isometries (ie Fig. \ref{fig5}(b)), we can heavily simplify the
structure of this contraction. For example, in the figure the tensors on the right
which are in a dimmer color can be substituted by straight lines, so that we
obtain the tensor contraction of Fig. \ref{fig6}(b). Now, by redrawing it we
obtain that of Fig. \ref{fig6}(c), which can in turn be written as a product
of matrices, as it was the case with MPS. When we have a product
of local observable, we can follow the same procedure. Just by having a look
at the diagram, we can eliminate some of the tensors and replace them by
lines, obtaining at the end a simple structure which can be easily contracted.

\begin{figure*}
\centering
\includegraphics[scale=0.7]{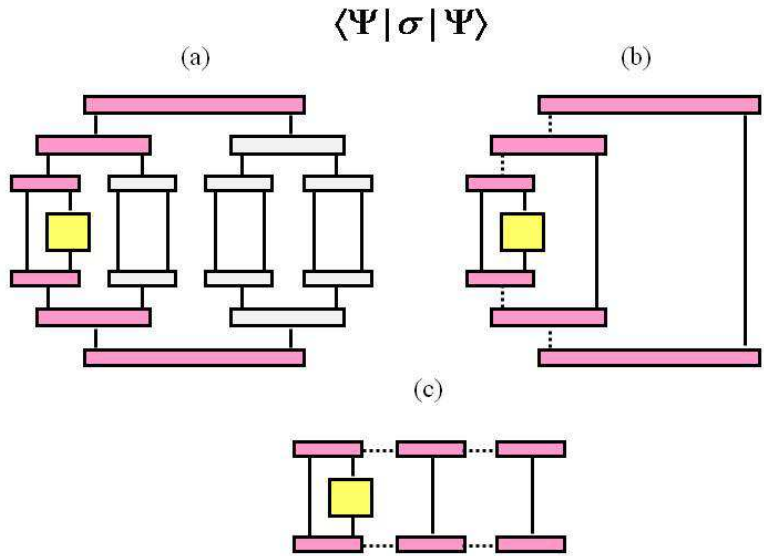} \\
\caption{\label{fig6}%
 Expectation values of observables in a TTS. (a) Contraction of the TTS with
 an observable and its conjugate. (b) Using the fact that the $T$'s are isometries
 we can get rid of several boxes (tensors); (c) reordering the indices we
 obtain a product of matrices.}
\end{figure*}

\subsection{Renormalization group}

As before, we can build a specific way of performing the renormalization, which is
nothing but the standard one (but in position space instead of momentum or energy
space). As in the case of the real-space renormalization group reviewed in previous
sections, the simplest method consist of trying to minimize the energy every time
we perform a renormalization step.

Let us consider a Hamiltonian of the form (\ref{Ham1}) (we could take other Hamiltonians
with longer range interactions, but for illustration purposes we take the simplest
one). First, we take $h_1$, acting on spins 1 and 2, and determine the subspace
of dimension $d_2^{\rm k}$ with lowest energy. That is, we diagonalize $h_1$ and take
the subspace spanned by the $d_2^{\rm k}$ lowest eigenvalues. The projector on that
subspace defined the isometry $T^{(2)}_1$. We do the same with $h_3$, $h_5$, etc.
Then, we project the whole Hamiltonian onto the subspace (which is build as a tensor
product of the selected ones), obtaining a new Hamiltonian with nearest neighbor
interactions only. The reason is that the projection of $h_{2M-1}$ is supported on
the subspace of the new particle $M$, whereas $h_{2M}$ is on that of particle $M$ and $M+1$
only. If we continue in this vein, we will obtain a renormalized Hamiltonian at each
step. The ground state of the final Hamiltonian will thus have the structure of
a TTS. Furthermore, this procedure may converge to a particular Hamiltonian, at least
if we consider translationally invariant systems, and which
is the fixed point operator of the renormalization group flow. This is the one
which one usually analyzes in renormalization group to determine the possible phases
depending on the different constants parametrizing $H$.

Of course, one may go beyond this procedure and try to minimize the energy directly
as a function of all the tensors appearing in the tree, in much the same way
one does in DMRG \cite{FriedmanTNS,MartinTNS,NagalTNS,ShiVidal2006}.
In fact, one can use similar techniques to those introduced
in that context in order to perform this task in an efficient way. For instance,
one may fix all the $T$'s except for one, and minimize the energy with respect to
that one. Then proceed with the next, and so on, until all the tensors (and thus
the energy) converge. One can also choose all the tensors in a row equal, in order
to emulate translationally invariance (this will occur anyway with the renormalization
procedure given above for Hamiltonians with that symmetry), at the expense that one has
to minimize with respect to all the tensors in one row at the same time. Furthermore,
one may look for quasi-scale invariant solutions, in which all the tensors are chosen
to be the same. Note, however, that unlike the case of MPS the
state is not translationally invariant, since the spins are treated on different
footing. That is, two spins that lie next to each other may not be grouped until
the final step of the renormalization procedure. This indicates that this method
may give good qualitative results (in most of the times, enough to determine the
phases appearing in the problem), although not as precise as in DMRG.

\subsection{Matrix Product States vs Tree Tensor States}

Given a MPS it is relatively simple to express it as a TTS. One way of seeing that is
by considering the sequential generation of MPS explained in previous sections. There,
we saw that a MPS of bond dimension $D$ can be generated by using a sequence
of unitary operators acting on $\log D+1$ spins (see Figs. \ref{fig4}, \ref{fig7}). If we group
$\log D$ spins into a single one, so that the new chain has $N/\log D$ big spins,
those unitary operations become between nearest neighbors only. Now, let us consider
the last of such spins. Obviously, there exists a unitary operation acting on
it and the previous one that disentangles it from the whole chain (such unitary
is the inverse of the one we apply in the sequential operation). The same applies
to any big spin, since we can consider it as the last one of a sequentially
generated state (we could start from the next one, go around the rest of
spins, and end up in that particular one). This implies that we can, for example,
act on every second big spin of the chain and its neighbor to the left with
a unitary operation and disentangle them completely from the rest (see Fig. \ref{fig7}).
The remaining state of the rest of the spins will still be a MPS with bond
dimension $D$, so that we can apply exactly the same procedure. By iterating,
we can completely disentangle all the big spins. The procedure we have carried
out is nothing else but the one described at the beginning of this section,
but with the big spins. This  means that the state
of the big spins can be written as a TTS, and
thus the original one too. The opposite is also true. Given a TTS we can always
use the procedure described in previous sections to write it as a MPS.
As discussed above in the context of the area law, the maximal rank of the density
operators $\rho_M$ will be $O(\log N)$, and thus this determines the largest bond dimension,
$D$, of the MPS.

\begin{figure*}
\centering
\includegraphics[scale=0.7]{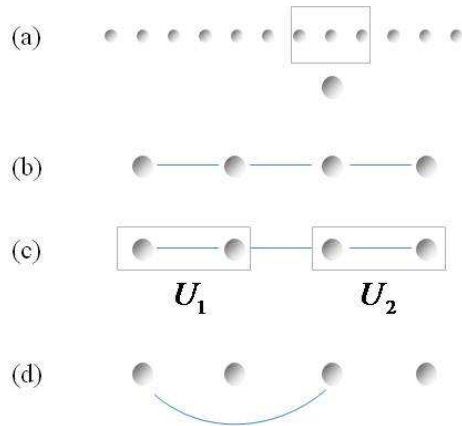} \\
\caption{\label{fig7}%
 A MPS can be expressed as a TTS. (a) We group $\log D$ spins to build
 bigger spins, (b), which are entangled and form an MPS of bond dimension
 $D$. (c) Unitary operators acting on pairs of spins disentangle half
 of them, and leave the rest in a MPS of bond dimension $D$. We can
 continue this procedure until we have a single spin.}
\end{figure*}

\subsection{Other remarks}

It is very clear that one can apply the same techniques described in the previous
section for MPS to TTS. For example, one can minimize the energy, determine the best
approximations, time evolution, etc, by getting the tensors $T$ of the TTS variationally,
going one by one (while fixing the rest), as in DMRG. The only point where one has to be careful is
that, in this case, the fact that each $T$ must be an isometry plays an
important role. In principle, the minimization of the energy, etc, for each
tensor should take this constraint into account. However, as it was in the
case of MPS, one can 'pull' this constraint so that in practice it is
irrelevant. The idea here is to perform the minimization in some particular
order. Let us take a position in the first row, fix the rest, and find
the optimal tensor at that position (without imposing the isometry condition).
Then, we determine a singular value
decomposition of the tensor between the upper index and the other two, keeping at the end
only the isometry. That is, we pull the non--isometric part upwards, and
include it in the next tensor. Thus, at the next step, we minimize with respect
of the tensor above the one before and continue in the same way until we
reach the uppermost position. Then, we go again to the first row and repeat the
whole procedure until we converge.
Apart from that, we can also define tree operators in analogy with matrix product
operators, and thus carry out calculations at finite temperature or with mixed states.

It is interesting to view what happens if we apply the renormalization group
procedure reviewed in this section to a finitely correlated state. In that case,
one can solve the problem exactly and classify the fixed points of that procedure,
obtaining the states that survive it \cite{VerstraeteLatorre05}.

\section{Multiscale Entanglement Renormalization Ansatz}

A more sophisticated way of implementing a real-space
renormalization group  was introduced by Ma and Dasgupta
\cite{MaDasgupta}, and later successfully used by Fisher
\cite{Fisher,Fisherb} in the context of random quantum spin systems. Their
blocking scheme is peculiar in the sense that one does not block
several spins into one superspin as described before, but maps $n$
spins into $n'<n$ spins, in such a way that locality in the
interactions is preserved. This constraint is of crucial importance
as it turns out that this is precisely the extra ingredient needed
to extend those block-transformation to higher dimensions such that
an area law for the entanglement entropy can be obtained. The idea of
mapping $n$ spins into $n'<n$ spins is also the basis of the multi-scale
entanglement renormalization procedure introduced independently by Vidal
\cite{vidal07a,vidal07b}.

To sketch this approach, let us consider a quantum spin chain with
only nearest neighbour interactions. A typical blocking scheme of
Ma-Dasgupta-Fisher in the case of Heisenberg interactions would map
4 nearest neighbor spin $1/2$'s $(k,k+1,k+2,k+3)$ into 2 new spin $1/2$'s
$(k',k'+1)$. This is done in such a way that the renormalized
Hamiltonian still only exhibits nearest neighbour interactions. In
other words, the isometry $U_{k',k'+1;k,k+1,k+2,k+3}$ used in the RG
step has the following effect on the Hamiltonian $H=\sum_k
h_{k,k+1}$:
\begin{eqnarray*}
&&U\left(h_{k-1,k}+h_{k,k+1}+h_{k+1,k+2}+h_{k+2,k+3}+h_{k+3,k+4}\right)
 U^\dagger\\
 &&=\bar{h}_{k-1,k'}+\tilde{h}_{k',k'+1}+\bar{h}_{k'+1,k+4}.
\end{eqnarray*}

Such RG steps can again be implemented recursively, and this has
been done with very big success for random antiferromagnetic spin
chains: the isometries $U$ are found using standard
second order perturbation theory, and the blocking becomes more
accurate as a function of the blocking step. In a similar vein, this
method has been used to simulate Ising spin systems with random
ferromagnetic interactions and random transverse fields \cite{Fisher,Fisherb}, in which
case 4 spins are mapped to 3 spins in such a way that the
renormalized Hamiltonian only exhibits nearest neighbour
interactions.

\begin{figure*}
\centering
\includegraphics[scale=0.7]{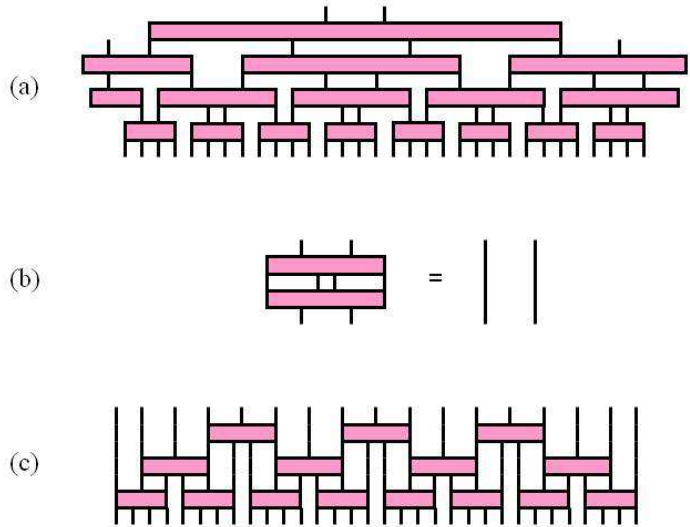} \\
\caption{\label{fig8}%
(a) Translational invariant blocking scheme a la
Ma-Dasgupta-Fisher mapping 4 spins to 2 by isometries. (b) Every
building block in the scheme consists of isometries. (c) Alternative
blocking scheme as e.g. used for the Ising model mapping 4 spins to
3.}
\end{figure*}

In analogy with the TTS, the quantum states that are generated
during such a blocking scheme can easily be represented using
isometries (see Fig. \ref{fig8}). One of the most interesting
features of the class of states generated during such a RG procedure
is the fact that they can be critical and scale-invariant. Indeed,
the Dasgupta-Ma-Fisher real-space RG method has been used to extract
critical exponents. By looking at the structure
of isometries, one indeed observes that the Schmidt rank when
cutting the chain in two halves can grow as the logarithm of the
size of the chain, as in the TTN: if one considers a periodic arrangement of the
isometries as in Fig. \ref{fig8}, a logarithmic number of layers of
blocking steps can contribute to generating entanglement. This has
to be contrasted to the case of MPS, where the Schmidt number with
respect to any cut is always bounded by a constant. Another very
important property of the states obtained like that is that one can
efficiently calculate expectation values of local operators: using
the RG-scheme, one can represent local operators in the effective
basis generated after consecutive RG-steps, and due to the
exponential shrinking of the number of spins at every step, local
operators will always remain local, up to the last level of the tree
where any expectation value can trivially be calculated. More
precisely, local operators will never act on more than a constant
number of renormalized spins during the renormalization flow.
This can be proven as follows: given an operator acting on
$x_k$ nearest neighbor spins after $k$ iterations, then there exist constants
$0<c_1<1$ and $c_2>1$ that depend on the RG blocking such that the
range of the operator at the next level is bounded by $x_{k+1}\leq
c_1\left(x_k+c_2\right)$. One can easily check that $x_n$ is always
bounded above by $\max(c1.c2/(1-c1),x_0)$ wich is a constant
independent of the number of spins. In the case of the $4\rightarrow
2$ isometry, $c_1=1/2,c_2=6$ and hence the bound is $\max(6,x_0)$;
for a general $n\rightarrow m$ scheme, $c_1=m/n, c_2=2.(n-1)$.

Obviously, this class of states encompasses the class of TTS, as the
latter is obtained in the special case of $n\rightarrow 1$ blocking
schemes. However, this new class of states also shares the lack of
translational symmetry
with the TTS, as opposed to the case of
MPS.

Now, a crucial step can be made in order to obtain a much
more powerful method. One can choose the isometries
at will, something which can lead to very different isometries than those obtained
by second order perturbation theory. The resulting renormalization
scheme is precisely the one introduced by G. Vidal and
the underlying states were called multiscale entanglement
renormalization ansatz (MERA)\cite{vidal07a,vidal07b}.
This class of states was introduced by G. Vidal,
and his construction was inspired by ideas originating in the field of quantum
information theory of how to parameterize states using quantum circuits. He also
proposed to obtain the isometries variationally. In a typical
realization of a MERA, the states are parameterized by
specifying a periodic pattern of isometries (the free parameters are
exactly the isometries) as shown in Fig \ref{fig9} and, furthermore,
the isometries are themselves decomposed into a sequence of
so-called disentangling unitaries and  isometries (see Fig
\ref{fig9}(b)). Apart from historical reasons, the incentive for
splitting the isometries into smaller building blocks is that this
allows for a more efficient calculation of local expectation values;
however, the disentangling unitaries are already implicitly present
in the Ma-Dasgupta-Fisher blocking scheme. There is however also a
very nice intuitive interpretation of the effect of those unitaries
as so--called disentanglers: before doing a blocking, those
disentanglers take care of removing entanglement within the block
with the outside.

\begin{figure*}
\centering
\includegraphics[scale=0.7]{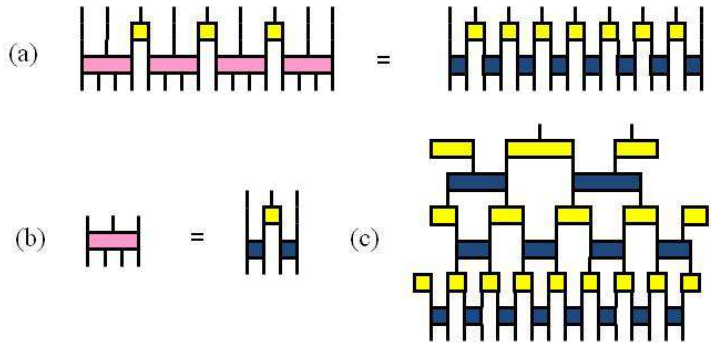} \\
\caption{\label{fig9}%
Multiscale Entanglement Renormalization Ansatz: (a),(b) Decomposing
the isometries in the Ma-Dasgupta-Fisher scheme into a sequence of
disentangling unitaries and isometries; (c) Typical representation
of a MERA as a sequence of unitaries and isometries.}
\end{figure*}

Due to the fact that MPS can be represented as TTS and TTS are a
special case of MERA, it is clear that the class of MERA encompasses
the class of MPS. As already explained in the context of the
Ma-Dasgupta-Fisher RG-scheme, expectation values of local
observables can easily be calculated by doing consecutive
coarse-graining steps on the obervable of the form
$\hat{O}\rightarrow U^\dagger\hat{O\otimes\one}U$. Due to the
exponential shrinking of the number of spins, it is guaranteed that
the renormalized observables $\hat{O}$ remain local at all steps.

In analogy to MPS and TTS, a variatonal calculation can now be done
as an alternating optimization over the degrees of freedom in the
state \cite{vidalmera2}. In the case of MERA, those degrees of freedom are the
isometries, but unlike the case of MPS and TTN, those optimizations
can not be mapped to alternating least squares problems; instead, a
direct optimization over isometries has to be done, which is a
nonlinear optimization problem that is more difficult and subtle to
control due to the occurrence of local minima. Nevertheless, Vidal
and collaborators have obtained impressive results, and e.g.
calculated critical exponents of quantum spin chains to a very good
precision by imposing a scale-invariant structure of the MERA \cite{vidalmera3,montangerocs}.
Such scale-invariant MERA have also been studied from the point of view of real-space
RG transformations \cite{Evenbly1} and of quantum channels and quantum
information theory \cite{Montangeromera,Montangeromerab}. A further development is the
formulation of a real- and imaginary-time evolution with MERA \cite{Rizzi1}.

\section{Higher dimensions}

In principle, one can use all the previous constructions in higher dimensions.
For instance, a MPS may approximate a 2D system, if we view it as a spin
chain (ie we place the spins one after each other)\cite{liang94,white96,xiang01}.
However, the validity of the methods explained in previous sections will
be questionable
as soon as the system becomes sufficiently big such that it cannot longer
be viewed as a 1D one. A way of expressing mathematically this intuition is
through the area law. One expects that the entropy of a region increases
with the number of spins at the border. For a MPS, it is simple to see
that there will be regions for which the the entropy will not scale at
all in that way. With TTS the same thing happens: whereas for some regions
the entropy will grow even as the logarithm of the number of spins on that
regions, for some other regions it will not grow at all. Apart from that,
they will not give a translationally invariant state, as it should be for
homogeneous problems. Nevertheless,
the renormalization scheme that originates the TTS may be still applied to
higher dimensional systems giving reasonable results for sufficiently small systems
\cite{VidalTNS}. The MERA can be chosen to fulfill the area law, and
thus they may be more appropriate than MPS and TTS for 2 and higher spatial
dimensions. The construction can be immediately adapted from the 1D one.
The issue of translational invariance still remains and
thus the result may depend on how the tensors are chosen.
Besides that, the minimization with respect to the tensors composing
the TPS cannot be carried out as efficiently as with TTN,
since it becomes difficult to avoid imposing the unitarity (or isometry)
condition on the tensors. In any case, the first results on a 2D frustrated
system in a 2D lattice have been recently reported, reveling a great
potentiality of the method \cite{VidalMERA09}.

Here we will mostly consider PEPS, which do
not suffer from some of those drawbacks, and for which it has been explicitly
proven to efficiently approximate a large set of problems, as mentioned in the introduction.
The prize one has to pay is that
the determination of expectation values has to be carried out approximately,
as opposed to what happens with MPS, TTS and MERA. In practice, this does
not pose a crucial problem since the error
in the approximation can be easily estimated and made arbitrarily small by
increasing the computational resources. PEPS algorithms have been recently
applied a variety of 2D problems with very promising results both
for finite \cite{verstraetecirac04,isacsson06,murgverstraete07,Murg09}
and infinite \cite{jordan07,orus09,Bela09} systems. We will also briefly
mention other states at the end of the section.

\subsection{Projeted Entangled Pair States in 2 dimensions}

We extend here the construction of MPS from a previous section to 2D \cite{verstraetecirac04}. For that,
we consider $N$ spins in a square lattice. We replace every spin by four
auxiliary ones (Fig. \ref{fig10}), each of them in a maximally entangled state
of dimension $D$ with a nearest neighbor (except for the borders). We then define
a map, $P_M$, acting on each of the sites, $M$, that transforms the state of the auxiliary
spins into the one of the original spins. Now, we can write the map at site
$M$ in a particular basis, so that
 \be
 P_M = \sum_{\alpha,\beta,\gamma,\delta=1}^D \sum_{n=1}^{d_1}
 (A_M^n)_{\alpha,\beta,\gamma,\delta} |n\rangle(\langle \alpha,\beta,\gamma,\delta|,
 \ee
where we have used the same notation as before, but now we have fixed the
dimension $D$ to simplify the notation. The state obtained after
this procedure is called a PEPS. We can write the PEPS in the spin bases,
in which case we will have that the corresponding coefficients will be given
by the contraction of the tensors $A$ according to the auxiliary indices (Greek
letters). This is represented in Fig. \ref{fig11}.

\begin{figure*}
\centering
\includegraphics[scale=0.7]{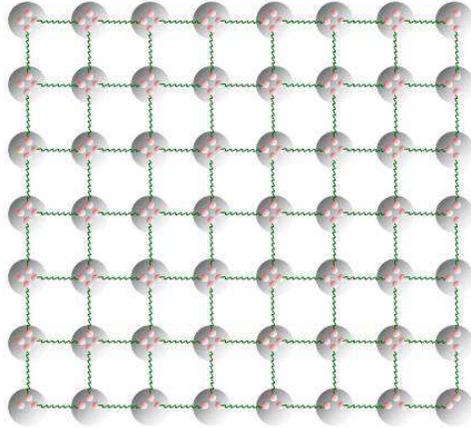} \\
\caption{\label{fig10}%
 Projected entangled pair state (PEPS). As in the 1D case, each
 spin is replaced by four ancillas, which are maximally entangled
 with their neighbors. The state is produced by locally mapping
 the states of the ancillas onto the original spins.}
\end{figure*}

Given the PEPS construction, it is very simple to understand that they satisfy the area law.
First, if we look at the state of the auxiliary particles, the entropy of
a region $A$ will be equal to the number of cuts, $n_A$, of the entangled pairs across
the border of the region times $\log D$. In fact, the rank of the reduced
density operator will be exactly $D^{n_A}$. On the other hand, the maps $P$ cannot
increase the rank of the density operator, and thus we obtain the area law
for the real spins, given that the entropy of an operator is upper bounded
by the logarithm of its rank.

\begin{figure*}
\centering
\includegraphics[scale=0.7]{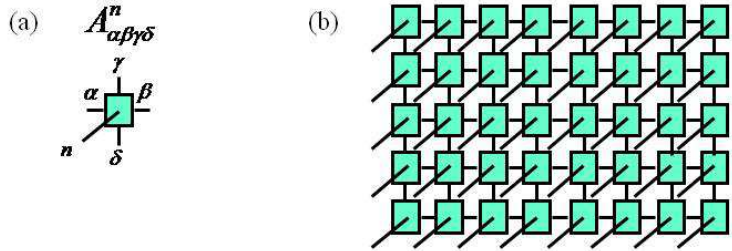} \\
\caption{\label{fig11}%
 Tensor network representation of a PEPS. (a) Representation of
 the tensor corresponding to a single site. The indices in
 the plane correspond to the auxiliary particles, whereas the one
 orthogonal is the spin one; (b) Representation
 of the whole state where the auxiliary indices are contracted. }
\end{figure*}

The expectation values of observables in a PEPS have a similar structure
to those in a MPS (see Fig. \ref{fig12}(a)). We have to sandwich the operator
between the tensors corresponding to $\Psi$ and $\bar\Psi$ as shown in
the figure. At the end, everything boils down to contracting a tensor
of the form shown in Fig. \ref{fig12}(b). This is very hard, in general.
The reason is that if we start contracting the tensors appearing there,
the indices will proliferate and in the middle of the calculation we will
have of the order of $\sqrt{N}$ indices, which amounts to having an exponential
number of coefficients. This is very different to what occurs in 1D, in which
chase the linear geometry makes it possible to contract the tensors while always
keeping two indices at most.

One way to proceed is to realize that the tensor network displayed in
Fig. \ref{fig12}(b) can be viewed as follows. The first row can be considered
as a tensor which in turns is built out of smaller tensors, in much the
same way as a MPS is built out of the tensors $A$. The next row can be viewed
as a MPO. Thus the contraction of the first row with the second will give
rise to a tensor like the first one but with a higher bond dimension (the square
of the original one). If we contract the next row, we will multiply the bond
dimension again, and at some point it will be impossible to proceed. Instead,
what we can do is to try to approximate the first plus second row by a tensor
of the form of a MPS but with reduced dimensions (Fig. \ref{fig12}(c)). For that, we use the technique
explained in previous sections on how to approximate states by MPS optimally.
We can now iterate this technique by adding the third row, and again decreasing
the bond dimension of the resulting tensor (Fig. \ref{fig12}(d)). In this way, we can keep all the
dimensions under control and obtain the desired expectation value.
A word of caution must be called for here. There is no guarantee that we will
be able to obtain a good approximation at the end if we decrease the dimensions
of the tensors. In practice, however, we have performed all those approximations
and obtained very good results. Furthermore, one is always aware of the error
made in the contraction, since it can be determined through the procedure itself.
The reason why one obtains very accurate results in practice can be qualitatively understood
as follows. The contraction we are performing can be viewed as evaluating a kind
of partition function of a 1D quantum system at non-zero temperature. In fact, the
MPO of each row can be interpreted as a transfer operator, in much the same
way as in 1D quantum (or 2D classical) systems. If that matrix has a gap, which
occurs outside the critical points, the procedure we are carrying out will tend to
give the eigenvector of the transfer matrix corresponding to the maximal eigenvalue.
If this eigenvector has an efficient representation as a MPS, then our procedure
will succeed. Even though there is no proof for that, the problem of finding the
maximal eigenvector of the transfer matrix in 1D is very reminiscent of that
of finding the ground state of a Hamiltonian, for which a MPS provides a good
approximation (for short range interactions, although numerically it also works
for longer range interactions).

Once one has an efficient algorithm to determine expectation values, one can
literally translate all the algorithms developed for MPS to PEPS. In particular,
one can approximate time evolutions, thermal states, etc, with this methods.
In Ref. \cite{reviewMurg} those algorithms are explained in great detail.

Let us now explain why PEPS are well suited to describe spins in thermal equilibrium in the case of
local Hamiltonians in any dimension. Let us write $H=\sum h_\lambda$, where
$h_\lambda$. For simplicity, we will assume that each $h_\lambda$ acts on
two neighboring spins although this can be generalized for $h_\lambda$ acting
on a small region.  We first rewrite the (unnormalized) density operator $e^{-\beta
H}=\tr_B[|\Psi\rangle\langle\Psi|]$, where $|\Psi\rangle=e^{-\beta H/2}\otimes|\Phi\rangle_{AB}$ is a
purification \cite{verstraeteripoll04} and $|\Phi\rangle_{AB}$ a pairwise maximally entangled state
of each spin with another one, the latter playing the role of an environment. We will show now that
$|\Psi\rangle$ can be expressed as a PEPS.  We consider first the simplest case where
$[h_\lambda,h_{\lambda'}]=0$, so that $|\Psi\rangle = \prod_{\lambda}e^{-\beta
h_{\lambda}/2}\otimes\one |\Phi\rangle_{AB}$. The action of each of the terms $e^{-\beta h_{\lambda}/2}$
on two spins in neighboring nodes can be viewed as follows
\cite{Cirac01,verstraetecirac04b}: we first include two auxiliary spins, one
in each node, in a maximally entangled state, and then we apply a local map in each of the nodes which
involves the real spin and the auxiliary spin, that ends up in $|0\rangle$. By proceeding in the same
way for each term $e^{-\beta h_{\lambda}/2}$, we end up with the PEPS
description. This is valid for all values of $\beta$, in particular for
$\beta\to\infty$, i.e., for the ground state. In case the local Hamiltonians do not commute, a more
sophisticated proof is required~\cite{hastings06}. One can, however, understand qualitatively why
the construction remains to be valid by using a Trotter decomposition to approximate $ e^{-\beta H}
\approx \prod_{m=1}^M\prod_\lambda e^{-\beta h_{\lambda}/2M}$ with $M\ll 1$. Again, this allows for a
direct implementation of each $\exp[-\beta h_\lambda/2M]$ using one entangled bond, yielding $M$ bonds
for each vertex of the lattice.  Since, however, the entanglement induced by each $\exp[-\beta
h_\lambda/2M]$ is very small, each of these bonds will only need to be weakly entangled, and the $M$
bonds can thus be well approximated by a maximally entangled state of low dimension. Note that the
spins belonging to the purification do not play any special role in this construction.

\begin{figure*}
\centering
\includegraphics[scale=0.7]{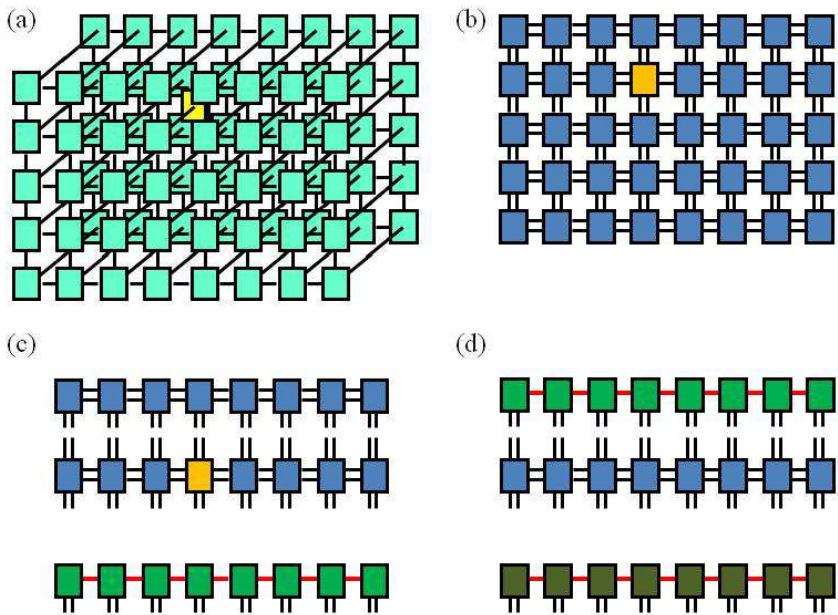} \\
\caption{\label{fig12}%
 Determination of expectation values with PEPS. (a) The tensor network
 corresponding to the expectation value is obtained by sandwiching the
 observable (here at the position 2,4) with $\Psi$ and $\bar \Psi$. (b)
 We can join pair of indices in each bond to have a single one. (c) In
 order to contract the resulting tensor, we observe that the first row
 has the same tensor structure as a MPS. Then, we contract the first row
 with the second, and approximate the result optimally by a set of tensors
 with the structure of a MPS. (d) We do the same thing with the next row,
 and continue in the same way until we find the result.}
\end{figure*}

\subsection{Other approaches}

For translationally invariant Hamiltonians, one may directly consider
the limit $N\to \infty$. In that case, the PEPS is taken with identical
maps $P$ (equivalently, tensors $A$), in as much the same way as MPS are
chosen to be FCS. The states so constructed have been
called iPEPS \cite{jordan07} and coincide
with the vertex--matrix product ansatz introduced earlier \cite{Sierra}.
In that paper, another family of translationally invariant states was introduced,
motivated by the interaction round the face models in statistical
mechanics. In a square lattice, the tensors $A$ are at the vertices of the B
sublattice. The physical index, $n$, of the tensor
$A^{n}_{\alpha_{11},\alpha_{12},\alpha_{21},\alpha_{22}}$ are associated
to the vertices of the B sublattice, whereas the
auxiliary indices, $\alpha_{ij}$, are at the bonds.
Thus, the contraction of the indices in that case
is different, since each auxiliary variable is common to four tensors
(and not two, like in the other case, see Fig. \ref{fig13}) (see also
\cite{Wolfgang,Rico}. One
can extend this last class of states to include other tensor structures
as shown in Fig. \ref{fig13}(c). In that state, there are two kinds of
tensors: those that are represented by circles and have one physical
index and four auxiliary ones, and those represented by squares with
no physical index \cite{Murgtobe}. The reason for the construction of those tensors
is that this is an effective way of increasing the
entanglement among nodes without increasing the bound dimension, but just the number
of tensors. Since the PEPS algorithms typically have a much milder dependence of
the computational time with the number of tensors than with the bond dimension,
this has a very positive effect on the algorithms. Finally, there is yet another
interesting class of states introduced by Nishino and collaborators, who
also extend the ones proposed by Sierra and Martin-Delgado as interaction round
a face type. They naturally appear in the transfer matrix of 3D classical models.
In those states, each tensor now belongs to a plaquette, and depends on all the
physical and auxiliary indices around the plaquette (ie. have the form
$A^{n_11,n_12,n_21,n_22}_{\alpha_{11},\alpha_{12},\alpha_{21},\alpha_{22}}))$.
Note that dropping the dependence on $n_{12},n_{21}$ and $n_{22}$ on that tensor
we obtain the one of Fig. \ref{fig13}(b).

\begin{figure*}
\centering
\includegraphics[scale=0.7]{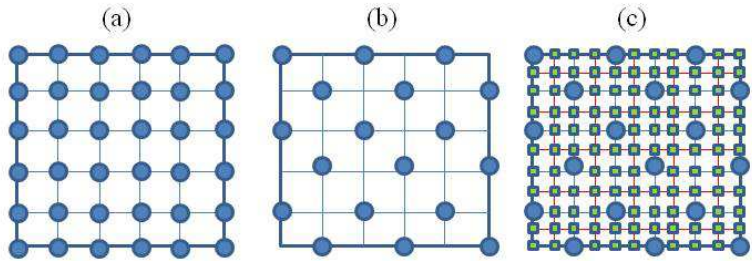} \\
\caption{\label{fig13}%
 Representation of a vertex (a) and a face (b)
 matrix product ansatz. The tensors are specified by circles, and the
 auxiliary indices by the lines joining them. As one can see, each
 auxiliary index appears in two (a) and four (b) tensors, respectively.}
\end{figure*}

We finally briefly mention other states which have an interesting property:
string-bond-states \cite{sbs} and the entangled-plaquette-states \cite{fabio}.
The coefficients of those states in the spin basis are expressed as products of
other coefficients. In the first case, the latter coefficients are just
MPS along different strings going through some spins in the lattice; in the second,
they correspond to large overlapping plaquettes. The main feature of those states
is that the expectation values can be calculated using Monte Carlo methods.
Other interesting states displaying that property have been introduced by
Sandvik \cite{sandvik08,sandvik08a,sandvikvidal}.

\section{Summary and perspectives}

It is very remarkable that one can identify the corner of Hilbert space which
is relevant to describe the ground and thermal equilibrium states for a large
variety of Hamiltonians. Those states are the MPS and their generalization
to higher spatial dimensions, the so-called PEPS. Here we have reviewed
those and other families of states (including the TTS and MERA) which are
a subclass of TPS. We have shown how most of them (all
except PEPS) follow immediately from some real-space renormalization procedure. We have
characterized them, and shown how one can build algorithms with them
to perform different tasks, ranging from finding ground states, thermal
states, evolutions, etc.

There remain very important problems in the context of TPS.
First of all, it would be very interesting to prove beyond the intuitive
arguments and the successful numerical results that the different procedures to determine expectation
values of PEPS converge in practice. Also, even though we know that the state
we are looking for is close to a MPS or a PEPS, nothing guarantees that our
algorithm (DMRG or its extensions) will find it, although in all practical
situations it does. Thus it would be very interesting to find conditions under
which this will be the case. Note that there exist problems for which the
ground state is a MPS but it cannot be found \cite{Schuchdiff}, or other in which it is a PEPS
and it cannot be efficiently contracted, since that would violate the general
believe in computer science that some problems are exponentially hard
\cite{verstraetewolf06,schuch07}.
Apart from that, when we have talked about the corner of Hilbert space we have
always restricted ourselves to ground or thermal states of short-range
interacting Hamiltonians. But, what happens for long-range Hamiltonians? Or,
for time evolution? In the last case it can be shown that even in 1D a MPS
can, in general, only approximate the state for short times
\cite{Schuchfaith07,CalabreseQuench}. This indicates
that the MPS are not well suited to describe time evolution for long times, and that the
family of states describing that corner of Hilbert space is a completely different one.

Another challenge is to find more efficient algorithms that work in higher
dimensions. For example, the time resources associated to the algorithm
based on PEPS to determine the ground state
of a 2D Hamiltonian with open boundary conditions scales like $N^2 D^{10}$, which
allows to work with up to $20\times 20$ lattices of spin 1/2 particles with
$D=5$. For problems with periodic boundary conditions or higher dimensions, the
dependence is even worse, which makes it unpractical. The same is true for
the MERA algorithms in 2 and higher dimensions. In particular, in 2D it scales
as $D^{16}$\cite{vidal2Dmera}. Thus, we have to find ways of
determining the states more efficiently, or new families of states (related to
PEPS) for which we can do this task much faster. One possibility which is currently
exploited and is very promising consists of combining the TPS
and other descriptions with Monte Carlo
methods \cite{sbs,fabio,sandvik08,sandvik08a,sandvikvidal}.

\ack
We wish to thank the people who has collaborated with us on the subject
of this manuscript. Special thanks go also to:
M. C. Banuls, J. von Delft, W. Duer, J. Garcia-Ripoll, M. Hastings,
C. Kraus, J. I. Latorre, I. Mc Cullogh, M. A. Martin-Delgado, F. Mezzacapo, V. Murg,
T. Nishino, R. Orus,
B. Paredes, D. Perez-Garcia,
B. Pirvu, D. Porras, E. Rico, M. Sanz, N. Schuch, U. Schollw{\"o}ck, G. Vidal, S. White,
and M. Wolf. This paper was written during a stay in the Aspen Center
for Physics, participating in the quantum simulation program. We
acknowledge financial support from the EU project QUEVADIS, the
German Science Foundation (FOR635), the excellence cluster
Munich Center for Advanced Photonics, and the FOQUS SFB from the
Austrian Science Foundation.

\end{document}